\documentclass[a4paper,fleqn,usenatbib]{mnras}
\usepackage{mathptmx}
\usepackage[T1]{fontenc}
\usepackage{ae,aecompl}
\usepackage{times}
\usepackage{amssymb}
\usepackage{amsmath}
\usepackage{upgreek}
\usepackage{color}
\usepackage{graphicx}
\usepackage{pdflscape}

\title[SCOPE Survey Description]{SCOPE: SCUBA-2 Continuum Observations of Pre-protostellar Evolution -- Survey Description and Compact Source Catalogue}
\author[D. J. Eden et al.]{D. J. Eden,$^{1,2}$\thanks{E-mail: D. J. Eden@ljmu.ac.uk} Tie Liu,$^{3,4}$ Kee-Tae Kim,$^{3}$ M. Juvela,$^{5}$ S.-Y. Liu,$^{6}$ K. Tatematsu,$^{7}$ \newauthor J. Di Francesco,$^{8,9}$ K. Wang,$^{10}$ Y. Wu,$^{11}$ M.A. Thompson,$^{2}$
% ...
\\
Affiliations are listed at the end of the paper}

\date{Accepted XXX. Received YYY; in original form ZZZ}

\pubyear{2018}

\begin{document}
\label{firstpage}
\pagerange{\pageref{firstpage}--\pageref{lastpage}}
\maketitle

\begin{abstract}

We present the  first release of the data and compact-source catalogue for the JCMT Large Program SCUBA-2 Continuum Observations of Pre-protostellar Evolution (SCOPE). SCOPE consists of 850-$\upmu$m continuum observations of 1235 \emph{Planck} Galactic Cold Clumps (PGCCs) made with the Submillimetre Common-User Bolometer Array 2 on the James Clerk Maxwell Telescope. These data are at an angular resolution of 14.4 arcsec, significantly improving upon the 353-GHz resolution of \emph{Planck} at 5 arcmin, and allowing for a catalogue of 3528 compact sources in 558 PGCCs. We find that the detected PGCCs have significant sub-structure, with 61 per cent of detected PGCCs having 3 or more compact sources, with filamentary structure also prevalent within the sample. A detection rate of 45 per cent is found across the survey, which is 95 per cent complete to \emph{Planck} column densities of $N_{\rmn{H_{2}}}$\,$>$\,5\,$\times$\,10$^{21}$\,cm$^{-2}$. By positionally associating the SCOPE compact sources with YSOs, the star formation efficiency, as measured by the ratio of luminosity to mass, in nearby clouds is found to be similar to that in the more distant Galactic Plane, with the column density distributions also indistinguishable from each other.

\end{abstract}


\section{Introduction}

The \emph{Planck} survey, with its primary goal of mapping the cosmic microwave background, covered the thermal emission from dust of $\sim$14\,K at wavelengths of 350\,$\upmu$m, 550\,$\upmu$m, and 850\,$\upmu$m. In the process of removing local, Galactic emission, a catalogue of 13188 \emph{Planck} Galactic cold clumps (PGCCs) was compiled \citep{Planck11,Planck16}. By comparing the column densities and velocity widths of such clumps and those containing active star formation, the PGCCs were found to be significantly more quiescent and less evolved \citep{Wu12,Liu13}. The apparent quiescent nature of PGCCs makes them a valuable sample for studying the earliest stages of star formation, especially since they appear to have conditions suitable for star formation, with low dust temperatures of 6--20\,K \citep{Planck16}. CO clumps have been detected towards PGCCs \citep[e.g.,][]{Liu13,Meng13,Parikka15,Zhang16,Feher17}, as well as detections of line emission from dense-gas tracers \citep{Yuan16}. There can be, however, low levels of active star formation within PGCCs \citep{Toth14,Liu15,Tang18,Yi18,Zhang18}.

A large sample of prestellar cores and clumps needs to be studied to understand the evolution of cores and clumps after the formation of a young stellar object (YSO). In support of the effort to collect a large sample of pre- and protostellar cores and clumps, we present SCOPE, the SCUBA-2 Continuum Observations of Pre-protostellar Evolution Large Program. The project aims to test the earliest stages of star formation by observing 1235 PGCCs with the wide-field sub-mm bolometer camera, the Submillimetre Common-User Bolometer Array 2 (SCUBA-2; \citealt{Holland13}) at the 850-$\upmu$m wavelength beam size of 14.4 arcsec on the James Clerk Maxwell Telescope (JCMT). By observing with SCUBA-2 in the 850-$\upmu$m continuum, matching the frequency of the \emph{Planck} 353\,GHz band, we can significantly improve over the 5$\arcmin$ resolution of the \emph{Planck} observations of PGCCs. The 5$\arcmin$ resolution of \emph{Planck} is prohibitive in that it hinders the positional cross matching of higher-resolution catalogues of YSOs \citep[e.g][]{Wright10,Gutermuth15,Marton16} and also doesn't reveal the highly structured nature of the PGCCs \citep{Juvela12,Liu12}.

The resolution of SCUBA-2 at 850\,$\upmu$m resolves PGCCs at 1.5\,kpc and 0.5\,kpc to scales of 0.1\,pc and 0.03\,pc, respectively, the typical size of star-forming cores and clumps \citep[e.g.,][]{Konyves15}, with 56 per cent and 43 per cent of sources falling within these distances, respectively \citep{Planck16}. It is important to resolve individual cores in the PGCCs, especially because of the close connection between the core mass spectrum and the the stellar IMF \citep[e.g.,][]{Simpson08,Konyves15,Montillaud15},

The SCOPE survey was awarded 300 hours in the JCMT weather bands 3 and 4, which correspond to sky opacity values of $\uptau_{225}$\,=\,0.08--0.2. Observations began with three periods of pilot observations occurring in September 2014, and continued from December 2015 until July 2017. SCUBA-2 does observe the 450-$\upmu$m band simultaneously with the 850-$\upmu$m band but the weather bands available to SCOPE do not allow for reliable photometric calibration to be made on these shorter wavelength data.

\subsection{SCOPE science goals}

As previously mentioned, the \emph{Planck} survey mapped the entire sky, therefore PGCCs cover all Galactic longitudes and latitudes, from the Galactic Plane to high-latitude clouds. Previous works suggest that the Galaxy's spiral arms do not have much of an impact on star formation, other than collecting the source material together \citep{Eden12,Eden13,Moore12}. Understanding star formation out of the Galactic Plane, however, is restricted to nearby clouds in the Gould's Belt \citep[e.g.,][]{Ward-Thompson07,Andre10}. Some clouds at higher latitudes show signs of star formation \citep[e.g.,][]{McGehee08,Malinen14,Kerp16}. Hence, high-latitude clouds could be contributing to Galactic star formation, both at present and in the future by providing source material for fresh star formation. Indeed, gas at high latitudes could be a part of a Galactic fountain, where gas and dust is expelled into the Galactic halo by supernovae and stellar winds \citep{Bregman80}. This material would then fall back to the Galactic Plane, cooling and condensing into atomic clouds, with observations detecting \ion{H}{i} high-latitude clouds \citep{Rohser16}. These clouds then fall back onto the Plane, replenishing the reservoir for star formation \citep{Putman12}. In this regard, these clouds would also contain some molecular material \citep{Magnani10,Rohser16a} and dust \citep{Planck11a}, therefore SCOPE is important in quantifying the amount of dense gas at high latitudes and to see if it has the same star formation efficiency as in the Plane.

Images of the dust-continuum emission found that the ISM and molecular clouds are highly filamentary. with the vast majority of clumps and cores lying on, or in, these structures \citep{Molinari10,Andre10,Andre14}. The formation mechanism of filaments is disputed, with global gravitational collapse \citep{Hartmann07}, large-scale colliding flows in the cloud-formation process \citep{Heitsch08}, and decaying supersonic turbulence \citep{Padoan07} being suggested as possible formation mechanisms. Observational evidence, however, has been unable to discriminate between these possibilities.  The fact that most cores, both pre- and proto-stellar, are found on these filaments implies that the filamentary structures are crucial to their formation, regardless of how the filaments form. Follow-up studies have also found PGCCs to be filamentary \citep[e.g.,][]{Rivera-Ingraham16,Kim17}, and the increased resolution of the SCOPE survey allows the determination of the detection rate of filaments and the fragmentation of filaments into clumps and cores. Such detection of filaments, and the placement of clumps/cores along them will allow for a greater understanding of the role that they play in star formation \citep{Liu17,Liu18,Juvela18}.

Using the method of \citet{Sadavoy13}, we can combine $\emph{Herschel Space Observatory}$ data and simultaneously derive the dust temperature, column density, and dust emissivity spectral index. By calculating these values, we can study how dust properties vary between different Galactic environments and between sources in different stages of the star-formation process, with some of the sources in this study already addressed by \citet{Juvela17,Juvela18}.

Multiple papers have already been published using the SCOPE data, namely the maps and images \citep{Liu16,Kim17,Tatematsu17,Liu17,Liu18,Juvela17,Tang18,Zhang18,Yi18,Juvela18}.

The layout of this paper is as follows: Section 2 introduces the observing strategy and complementary observations and surveys. Section 3 describes the data and the data reduction, whilst Section 4 includes the compact source extraction. Sections 5 and 6 describe data access and results, respectively. Finally, Section 8 provides a summary and final conclusions.

\section{SCOPE Observing Strategy}

The SCOPE selection of PGCCs was chosen randomly from the full PGCC sample to sample varying Galactic environments \citep{Liu17}. Sources from the catalogue of PGCCs were excluded if they had already been observed at the JCMT with SCUBA-2 at a rms sensitivty of at least 6\,mJy\,beam$^{-1}$. These observations formed part of the JCMT Legacy Surveys \citep{Chrysostomou10}, namely the JCMT Plane Survey (JPS; \citealt{Moore15,Eden17} and the JCMT Gould Belt Survey (GBS; \citealt{Ward-Thompson07}). The SCUBA-2 Ambitious Sky Survey (SASSy; \citealt{MacKenzie11,Nettke17}; Thompson et al., in preparation), covers half of the Galactic Plane but does not have the desired rms.

The PGCCs not previously observed at JCMT were placed into a 3-dimensional grid with each longitude bin 30 degrees wide, latitude bins of $\mid$\emph{b}$\mid$\,=\,0$\degr$, 4$\degr$, 10$\degr$, and 90$\degr$, and distance bins of 0\,kpc, 0.2\,kpc, 0.5\,kpc, 1\,kpc, 2\,kpc, and 8\,kpc. Sources with column densities N$_{\rmn{H_{2}}}$ $<$ 2\,$\times$$10^{21}$ cm$^{-2}$ chosen randomly within each bin, with all sources with column densities N$_{\rmn{H_{2}}}$ $>$ 2\,$\times$$10^{21}$ cm$^{-2}$ observed. This column density criterium is due to the pilot study of the initial 300 PGCCs, where the detection rate of PGCCs with N$_{\rmn{H_{2}}}$ $<$ $10^{21}$ cm$^{-2}$ were dramatically lower. The JCMT has through SCOPE, and other archival data, covered nearly all PGCCs with column densities of N$_{\rmn{H_{2}}}$ $>$ 2\,$\times$$10^{21}$ cm$^{-2}$  within the declination range of -30$\degr$ -- +35$\degr$, the most likely star-forming PGCCs (Wang et al., in preparation).

\begin{figure*}
\begin{tabular}{lll}
\includegraphics[width=0.32\linewidth]{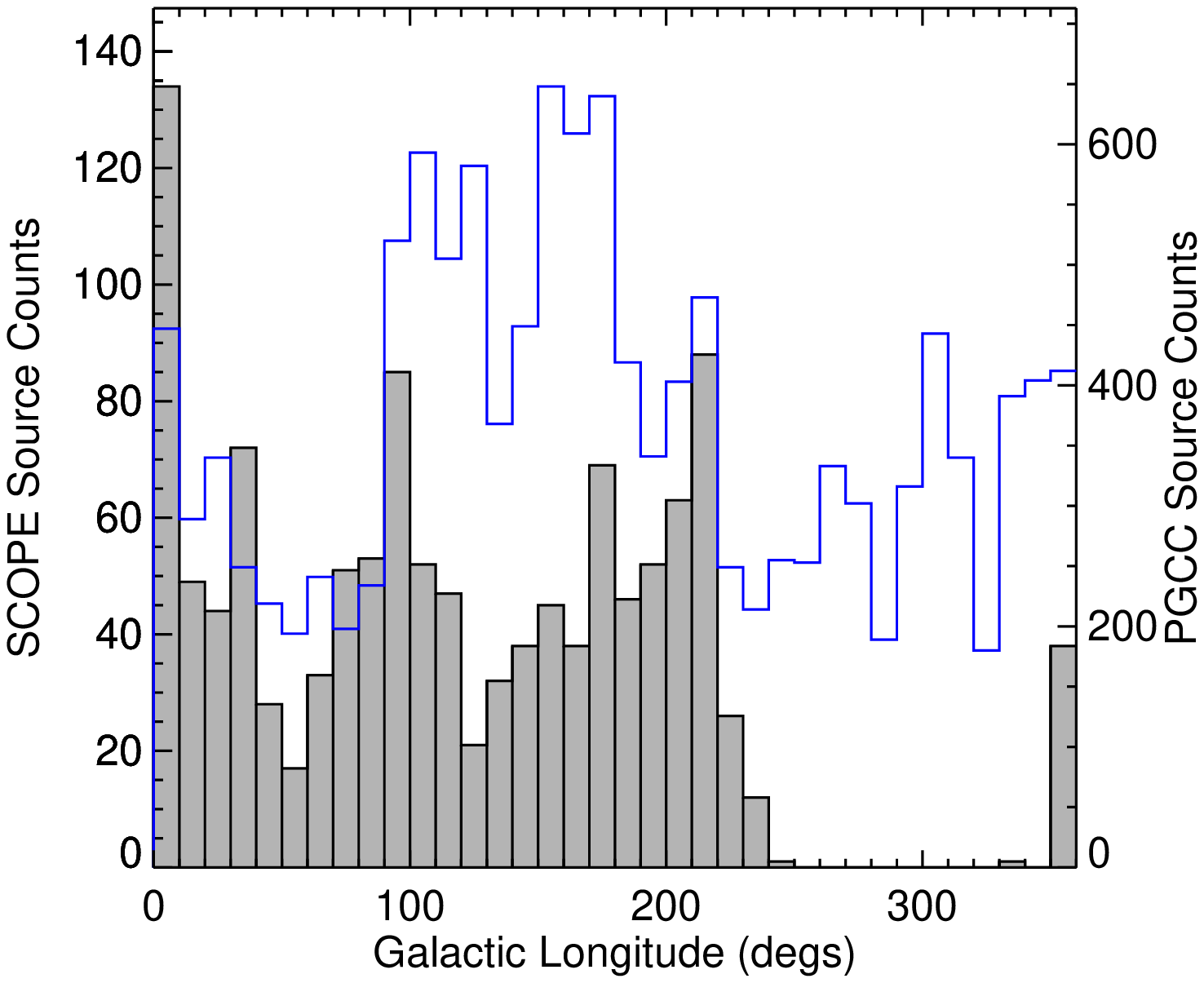} & \includegraphics[width=0.32\linewidth]{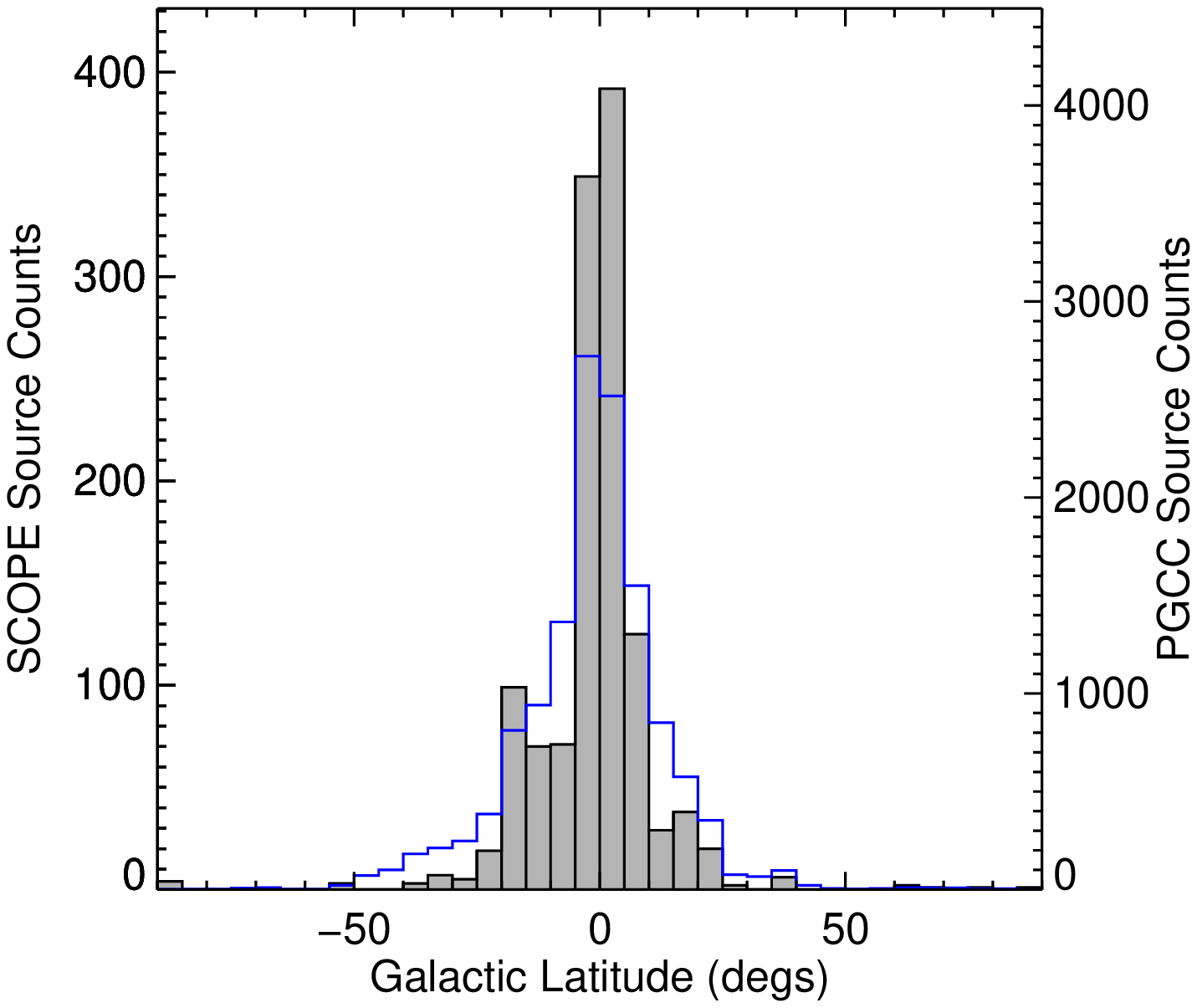} & \includegraphics[width=0.32\linewidth]{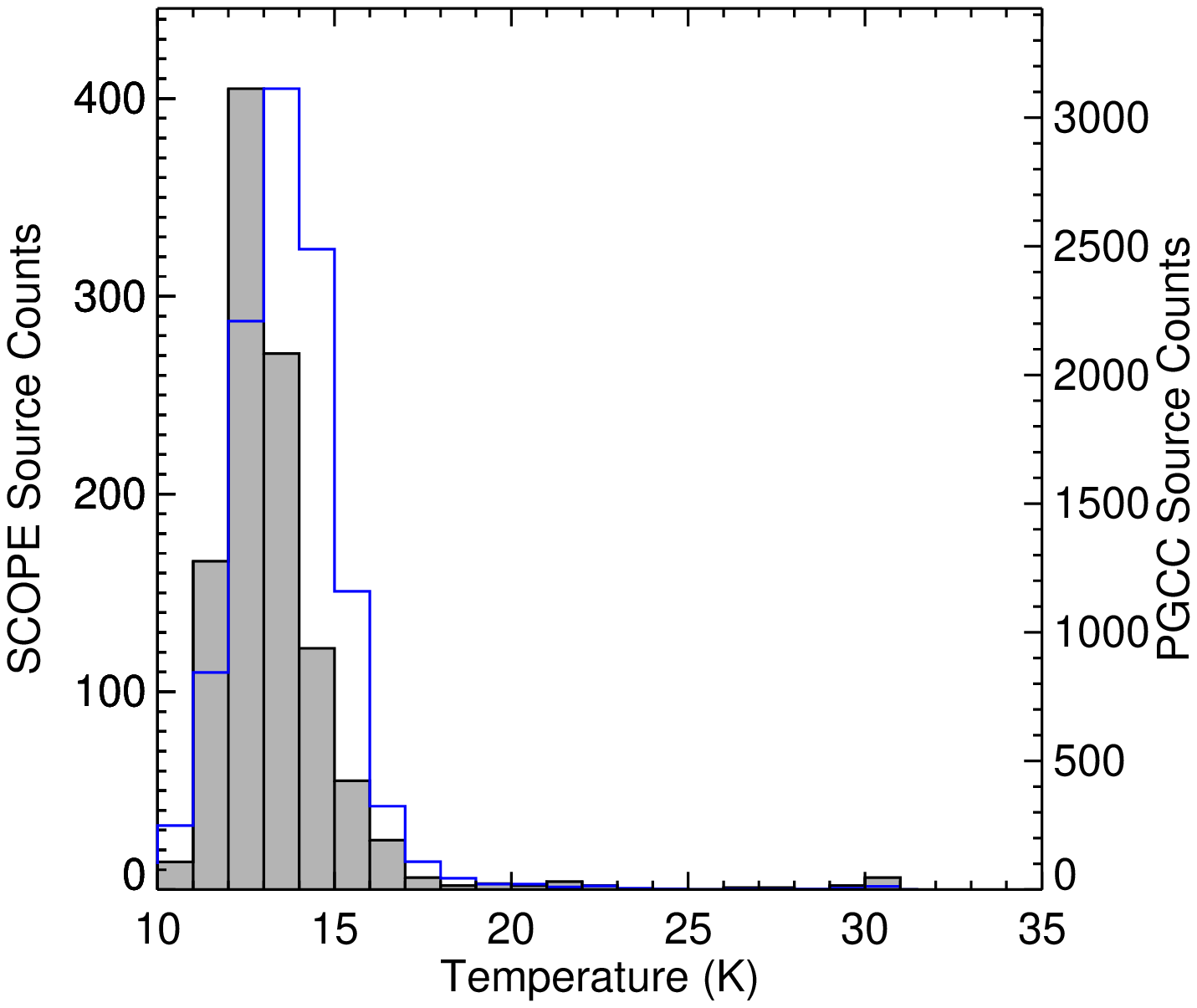}\\
\includegraphics[width=0.32\linewidth]{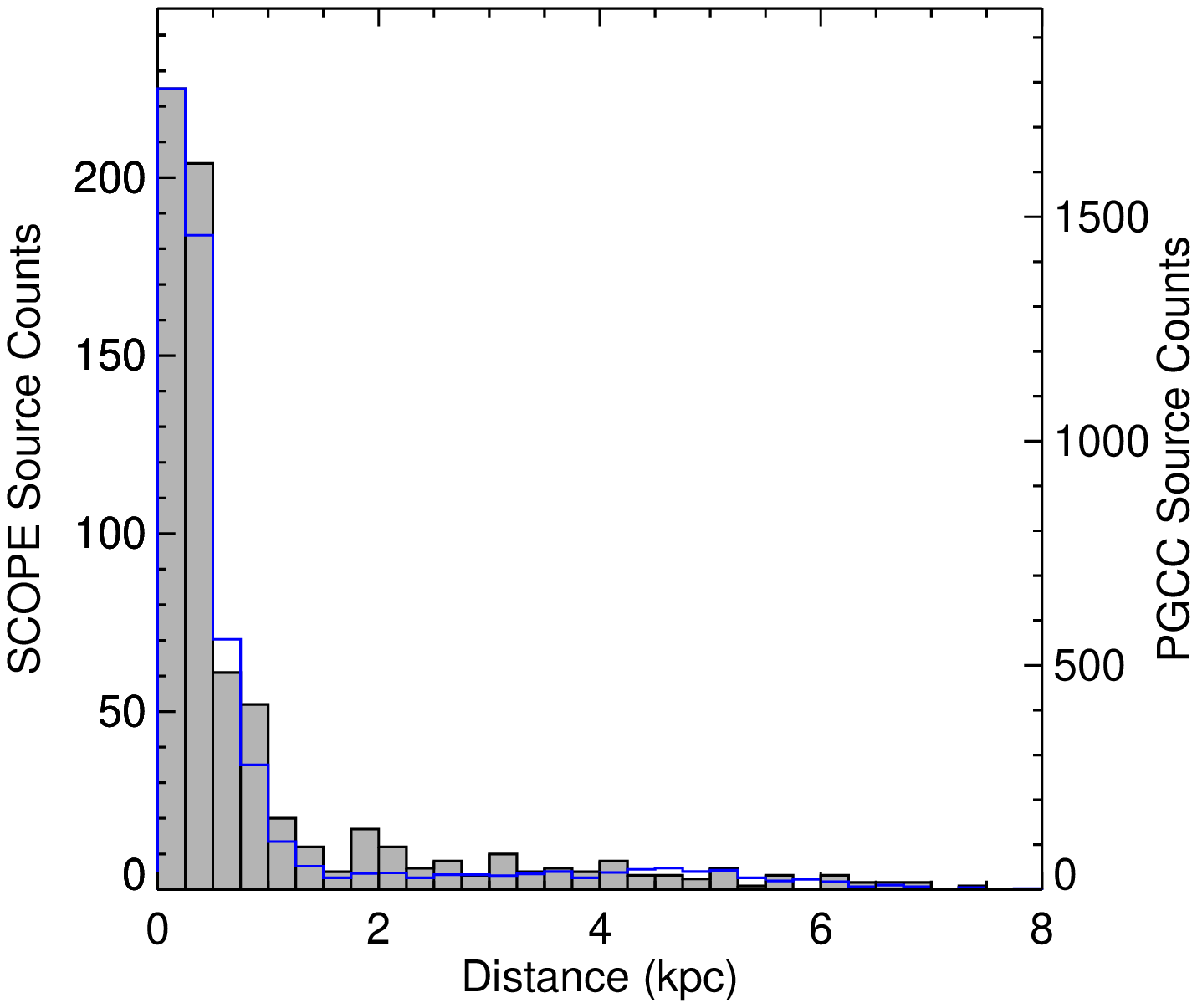} & \includegraphics[width=0.32\linewidth]{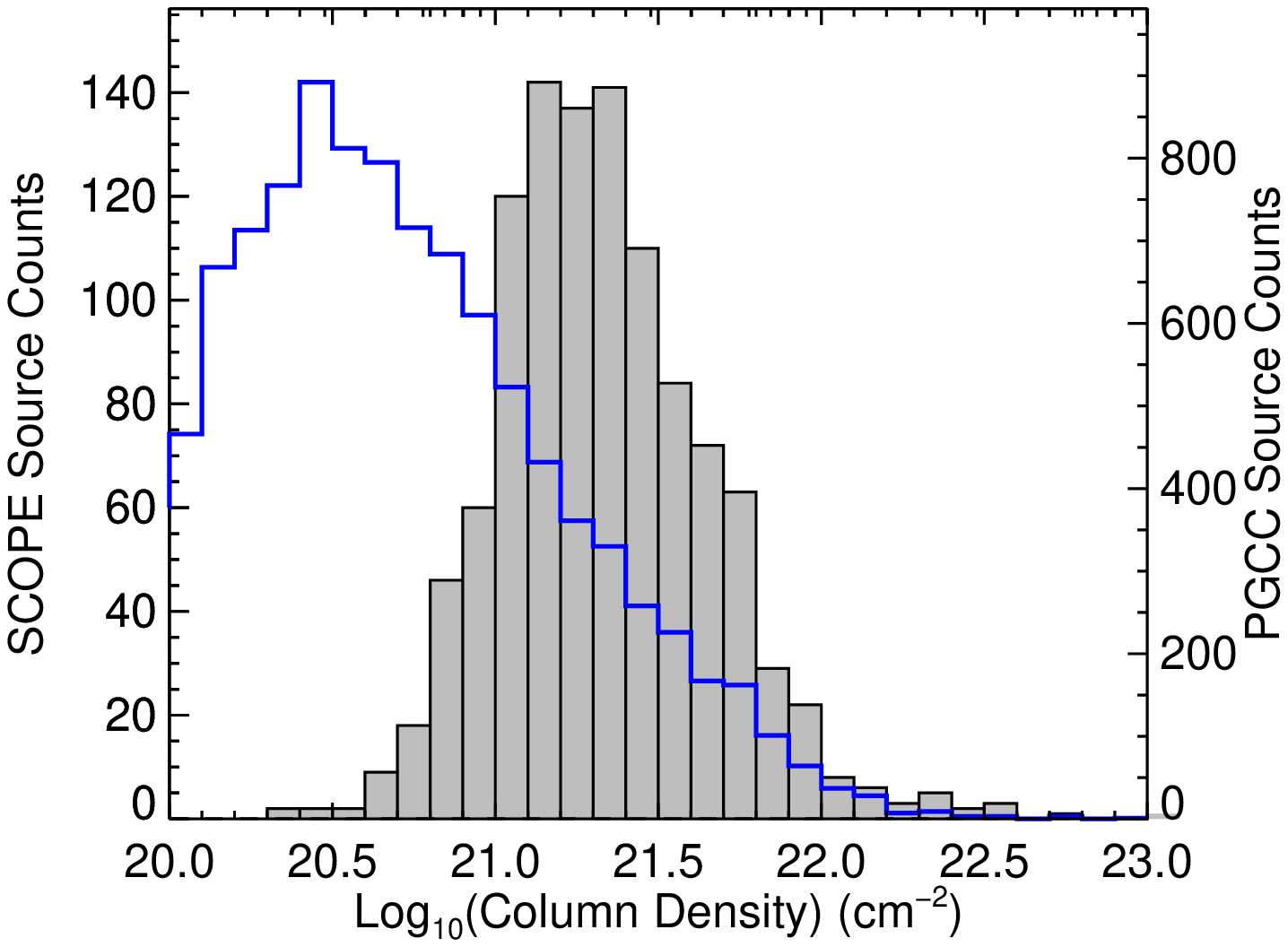} & \includegraphics[width=0.32\linewidth]{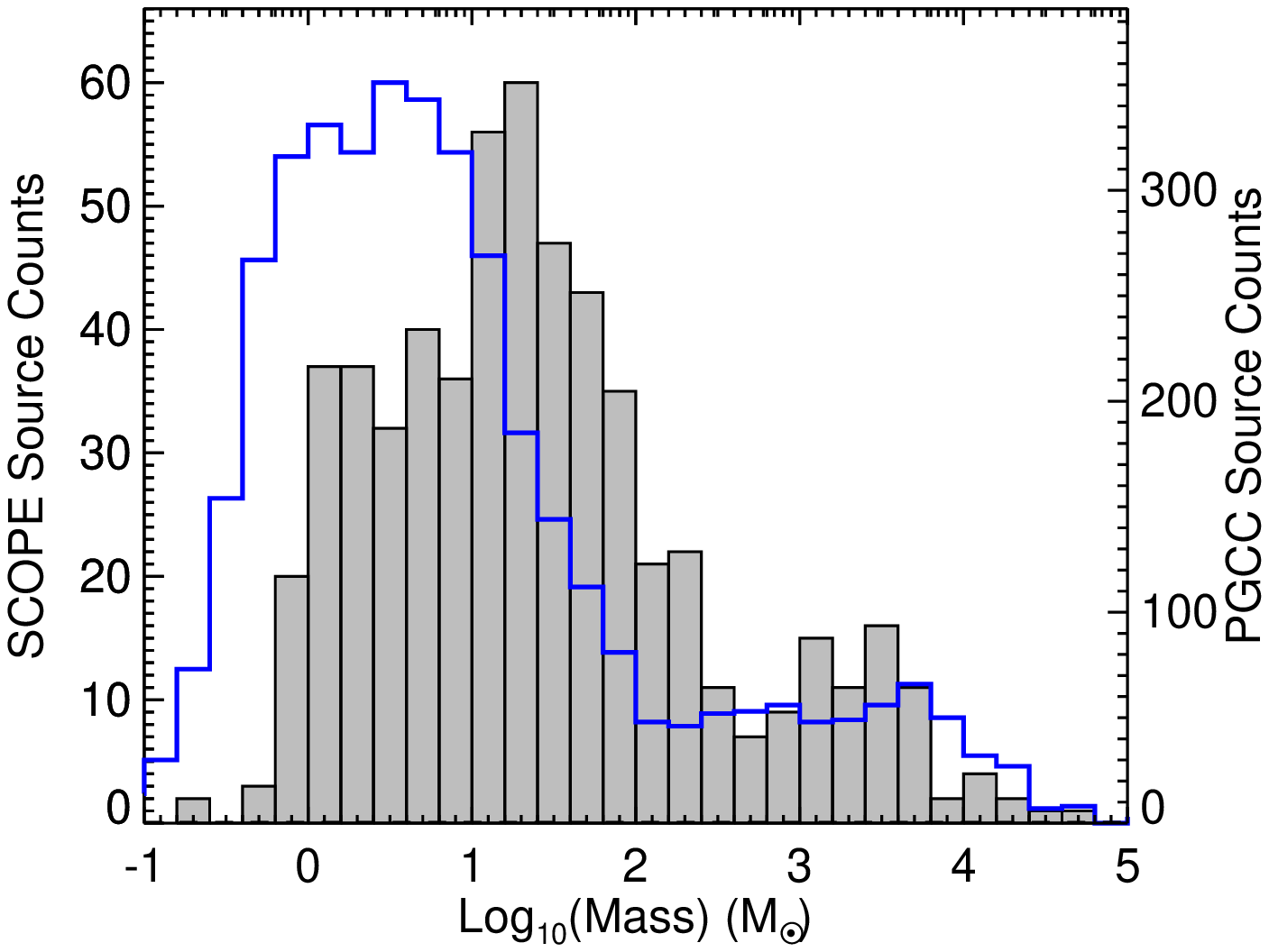}\\
\includegraphics[width=0.32\linewidth]{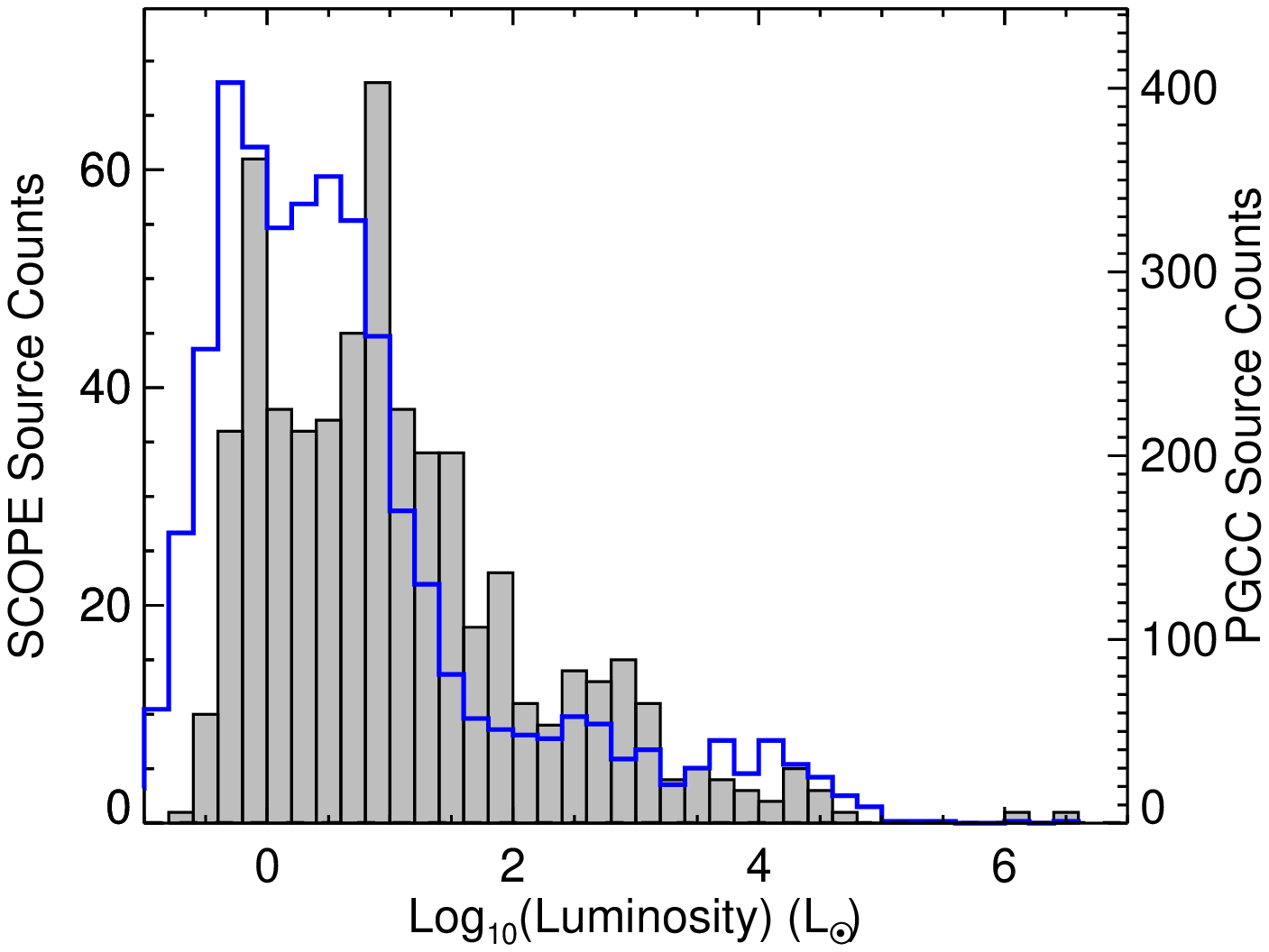} & \includegraphics[width=0.32\linewidth]{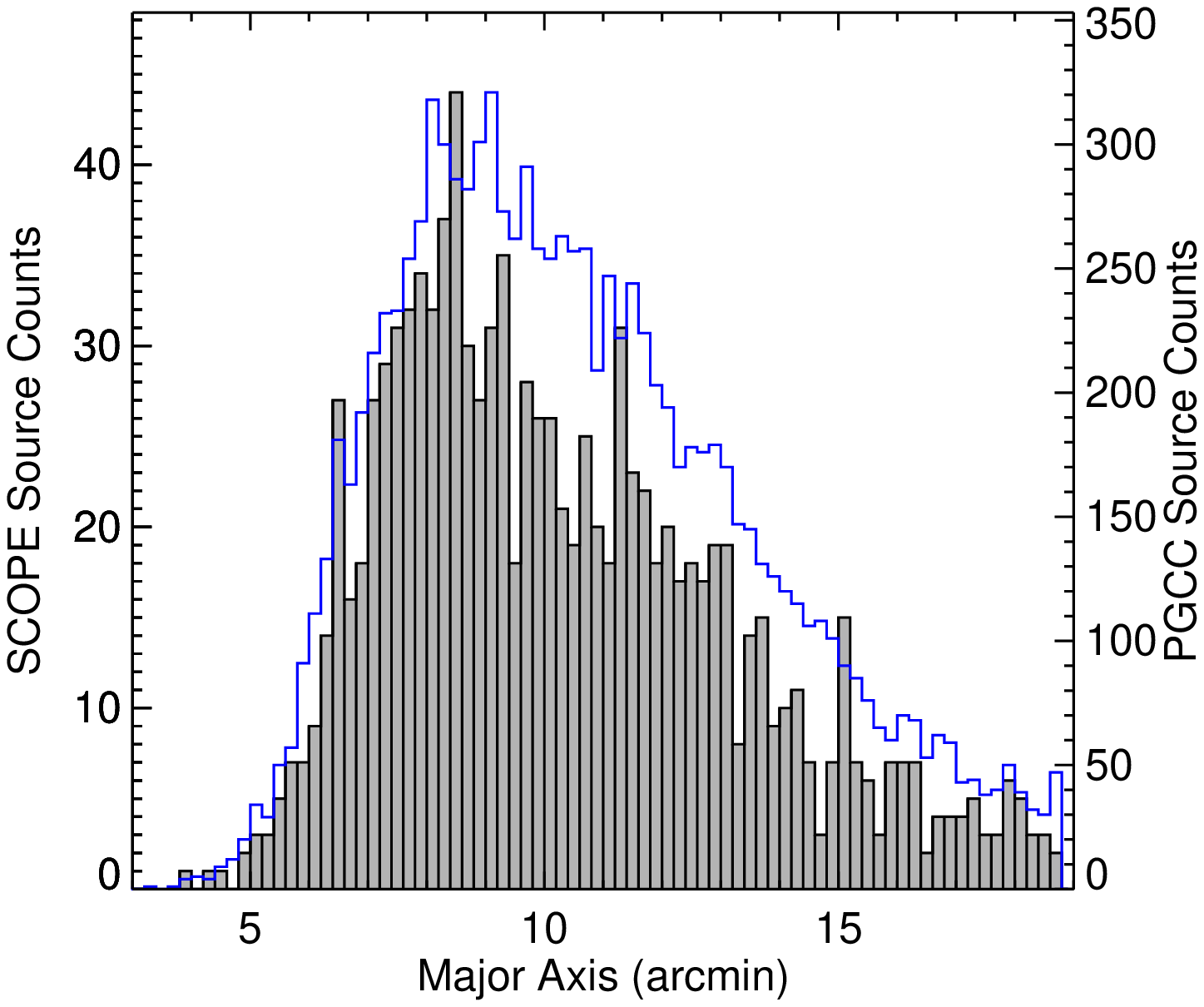} & \includegraphics[width=0.32\linewidth]{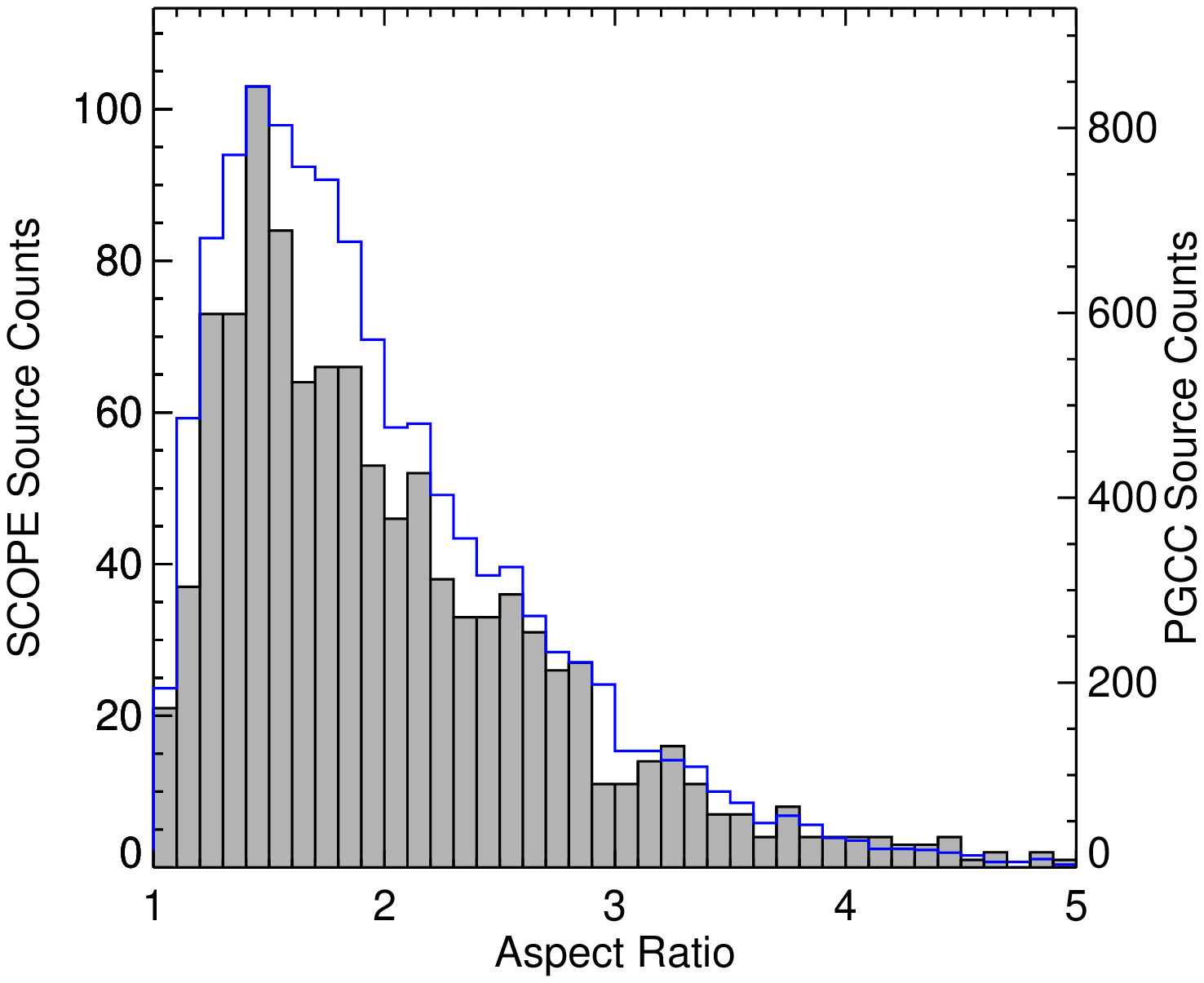}\\
\end{tabular}
\caption{Properties of the PGCCs observed in the SCOPE survey (grey histogram), with the total PGCC population from \citet{Planck16} overlaid in blue. The top row displays the Galactic longitude, Galactic latitude, and temperature, in the left, central, and right panels, respectively. The middle row are the distance, column density, and mass, with the bottom row showing the luminosity, major axis, and aspect ratio.}
\label{statistics}
\end{figure*}

The properties of the observed sources are displayed in Fig.~\ref{statistics}. The presented properties are Galactic longitude, Galactic latitude, temperature, distance, column density, mass, luminosity, major axis, and aspect ratio. The physical properties are not derived for all sources, and are as calculated by \citet{Planck16}. These distributions are compared to the entire populations, as derived by \citet{Planck16}. The biggest departure is with column density. We observed higher column density PGCCs than the average, which in turn skewed the SCOPE mass distribution to higher observed masses. Low column density (N$_{\rmn{H_{2}}}$$<$10$^{21}$ cm$^{-2}$) PGCCs usually do not contain dense cores, based on our pilot observations, and thus are not interesting for star formation studies.

The average angular size of PGCCs is 8$\arcmin$ \citep{Planck16}, therefore the CV Daisy mode of SCUBA-2 was used for these observations \citep{Bintley14}. The CV Daisy mode is specifically designed for small, compact objects, with the telescope keeping to a circular pattern at 155\,$\arcsec$\,s$^{-1}$. Each SCOPE map takes 16 minutes to perform, with the CV Daisy mode producing 850-$\upmu$m rms noise values of 6\,mJy\,beam$^{-1}$ in the central 3$\arcmin$, with 850-$\upmu$m rms noise out to radii of 6$\arcmin$ of 10-30\,mJy\,beam$^{-1}$.

\begin{figure*}
\includegraphics[width=\linewidth]{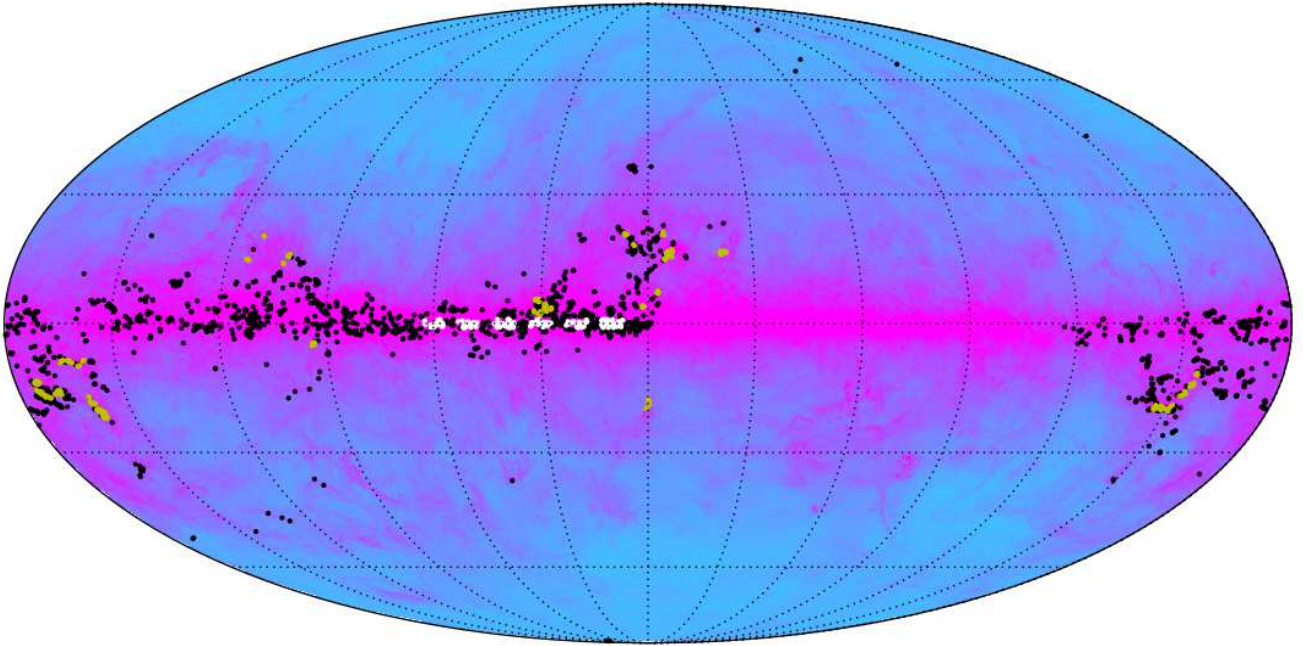}
\caption{The distribution of observed PGCCs by the JCMT. The black points represent SCOPE sources, whilst the white and gold points were observed as part of the JPS and GBS, respectively. The underlying image is the \emph{Planck} 353-GHz (850-$\upmu$m) intensity map.}
\label{allsky}
\end{figure*}

The SCOPE survey including pilot studies observed fields towards 1235 PGCCs in 1062 fields, with compact source detections towards $\sim$ 51 per cent of fields, 45 per cent of PGCCs. The observed PGCCs in the SCOPE survey are displayed in their Galactic context in Fig.~\ref{allsky}, with the 174 and 321 PGCCs observed by the JPS and GBS, respectively, overlaid. This paper will deal with only those observed as part of the SCOPE project.

\begin{figure}
\includegraphics[scale=0.5]{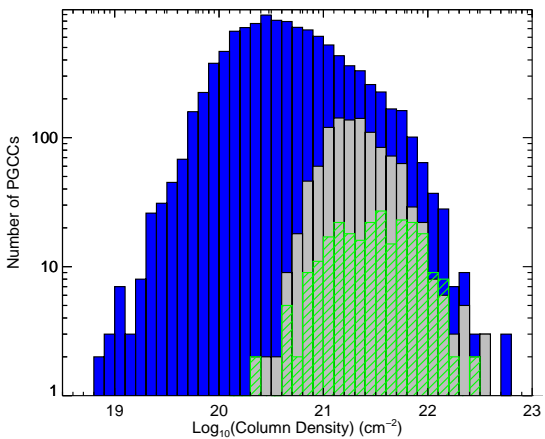}
\caption{Histograms of column density of PGCCs with the entire population represented by the blue histogram, the grey histogram representing those PGCCs observed by SCOPE and the green hashed histogram covering those PGCCs observed by the JCMT in the Gould's Belt Survey (\citep{Ward-Thompson07}.}
\label{gouldsbelt}
\end{figure}

The goal of SCOPE was to observe PGCCs in all Galactic environments. Comparison to the entire PGCC population and to sources in the GBS fields (Fig.~\ref{gouldsbelt}) shows that SCOPE observations are representative of the high-column-density PGCC targets and complement the sources observed in GBS.

\subsection{Complementary Observations}

As well as the SCOPE survey, there are ongoing surveys at other facilities, giving complementary data to these PGCC detections \citep{Liu15,Liu17}. The Taeduk Radio Astronomy Observatory (TRAO) Observations of \emph{Planck} cold clumps (TOP) survey is observing $\sim$\,2000 PGCCs in the rotational transition of $J\,=\,1-0$ of CO istopologues $^{12}$CO and $^{13}$CO at resolutions of 45-47 arcsec. A full description of this survey can be found in \citet{Liu17}. Further $J\,=\,1-0$ observations have been made at the Purple Mountain Observatory (PMO) at the same resolution. These observations will allow PGCCs to be put in the greater context of extended CO emission and structure.

The SMT (Submillimetre Telescope) ``All-sky'' Mapping of \emph{Planck} Interstellar Nebulae in the Galaxy (SAMPLING; \citealt{Wang18}) survey follows up PGCCs in the $J\,=\,2-1$ transition of $^{12}$CO and $^{13}$CO. The detection of two transitions will allow more accurate column densities to be calculated.

Further PI observations have occurred at the Nobeyama Radio Observatory 45-metre telescope, the 21-metre telescopes in Korean VLBI Network (KVN) and the Effelsberg 100-metre telescope in the dense-gas tracers like HCO$^+$, N$_{2}$H$^{+}$, HC$_{3}$N, CCS, DNC, HN$^{13}$C, N$_{2}$D$^{+}$ and NH$_{3}$ towards samples of SCUBA-2 dense cores in PGCCs. By observing these species, temperatures, depletion and deuteration fractions can be determined \citep{Tatematsu17}.

\section{Data Reduction and SCOPE Data}

\citet{Liu17} provides a full description of different data reduction methods employed and tested within the SCOPE survey.

The data reduction employed within the first SCOPE data release use the Dynamic Iterative Map-Maker \citep{Chapin13}, part of the $\emph{Starlink}$ {\sc smurf} package \citep{Jenness11}. These data make use of a 200$\arcsec$ spatial filter, with no use of external masking. This initial data reduction is then filtered for regions with a signal-to-noise ratio (SNR) less than 3, with these high-SNR regions used then as a mask for a further reduction. A full description of the masking process can be found in \citet{Mairs15}.

A flux conversion factor of 554 Jy\,beam$^{-1}$\,pW$^{-1}$ is used to convert from the native units of pW to Jy\,beam$^{-1}$. This value is $\sim$\,3 per cent higher than the 537 Jy\,beam$^{-1}$\,pW$^{-1}$ flux conversion factor recommended by \citet{Dempsey13}, reflecting the pixel size (4$\arcsec$) and data reduction method used by the SCOPE survey as the pixel size is a factor in the flux conversion factor equation.

Examples of two of the observed PGCCs are shown in Fig.~\ref{examples}, a complex, filamentary source, and a high-latitude cloud with simple morphology.

The mean rms within the central 12$\arcmin$ of the 1062 fields (this is the number of fields observed, containing the observed 1235 PGCCs) is 0.185 Jy\,arcsec$^{-2}$, which corresponds to 43.9\,mJy\,beam$^{-1}$. This sensitivity is a factor of $\sim$\,1.5 worse than the 25-31\,mJy\,beam$^{-1}$ rms of the JPS \citep{Eden17}. Within the central 3$\arcmin$, the rms is found to be 0.028 Jy\,arcsec$^{-2}$, which corresponds to 6.65\,mJy\,beam$^{-1}$.

\section{Compact Source Catalogue}

\subsection{Compact Source Extraction}

Compact sources were extracted from the images using the {\sc FellWalker} algorithm (FW; \citealt{Berry15})\footnote{FW is part of the $\emph{Starlink}$ {\sc cupid} package outlined in \citet{Berry07}.}. Justification for the choice of source extraction algorithm and the parameters used are outlined in \citet{Moore15} and \citet{Eden17}. The parameter {\sc FellWalker:minpix} is adjusted to account for the larger pixel size used in the SCOPE reduction.

The FW algorithm is run on the SNR maps, with the mask produced by {\sc cupid:findclumps} used to extract flux from the emission maps.

\begin{figure*}
\begin{tabular}{ll}
\includegraphics[width=0.49\linewidth]{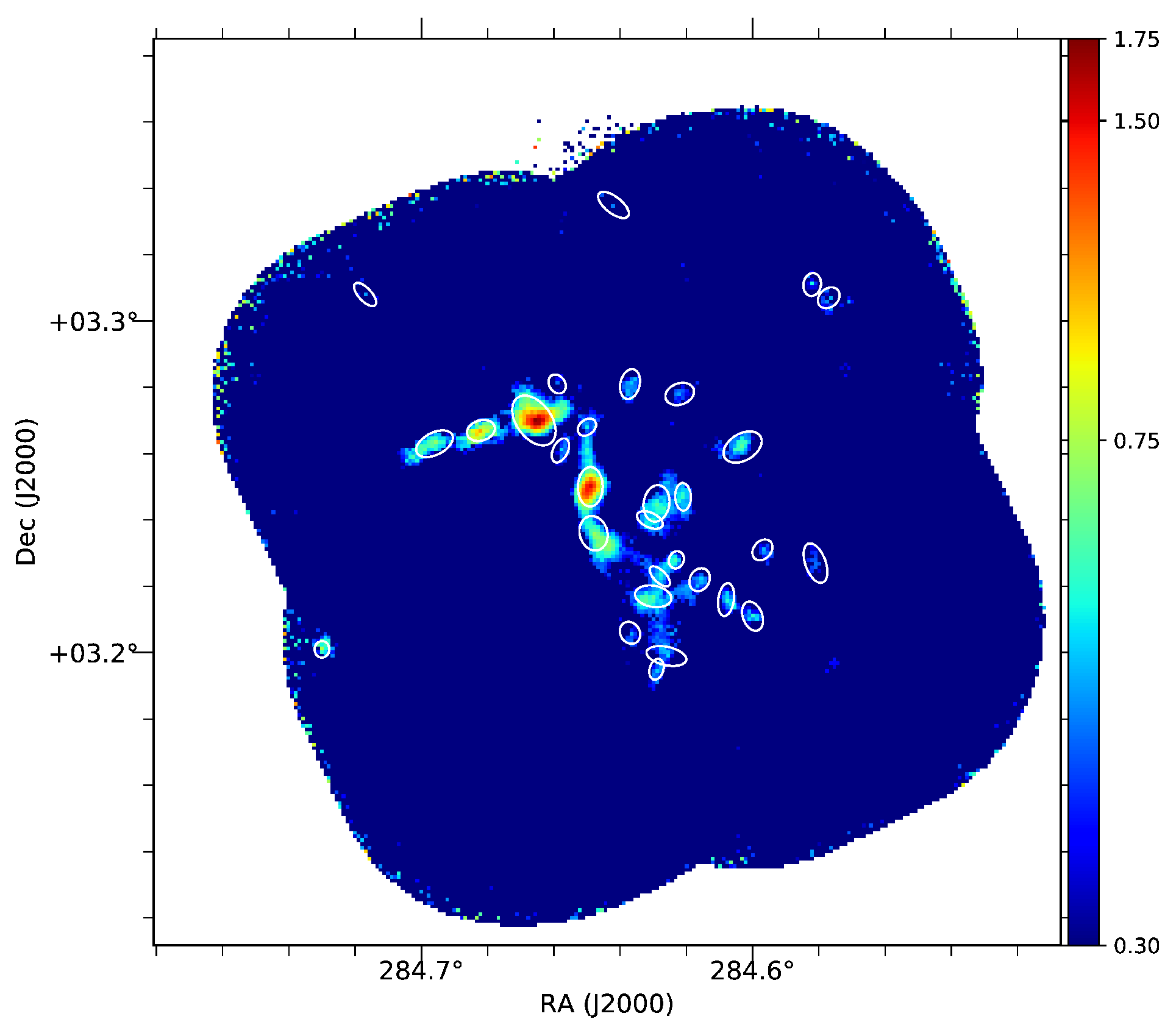} & \includegraphics[width=0.49\linewidth]{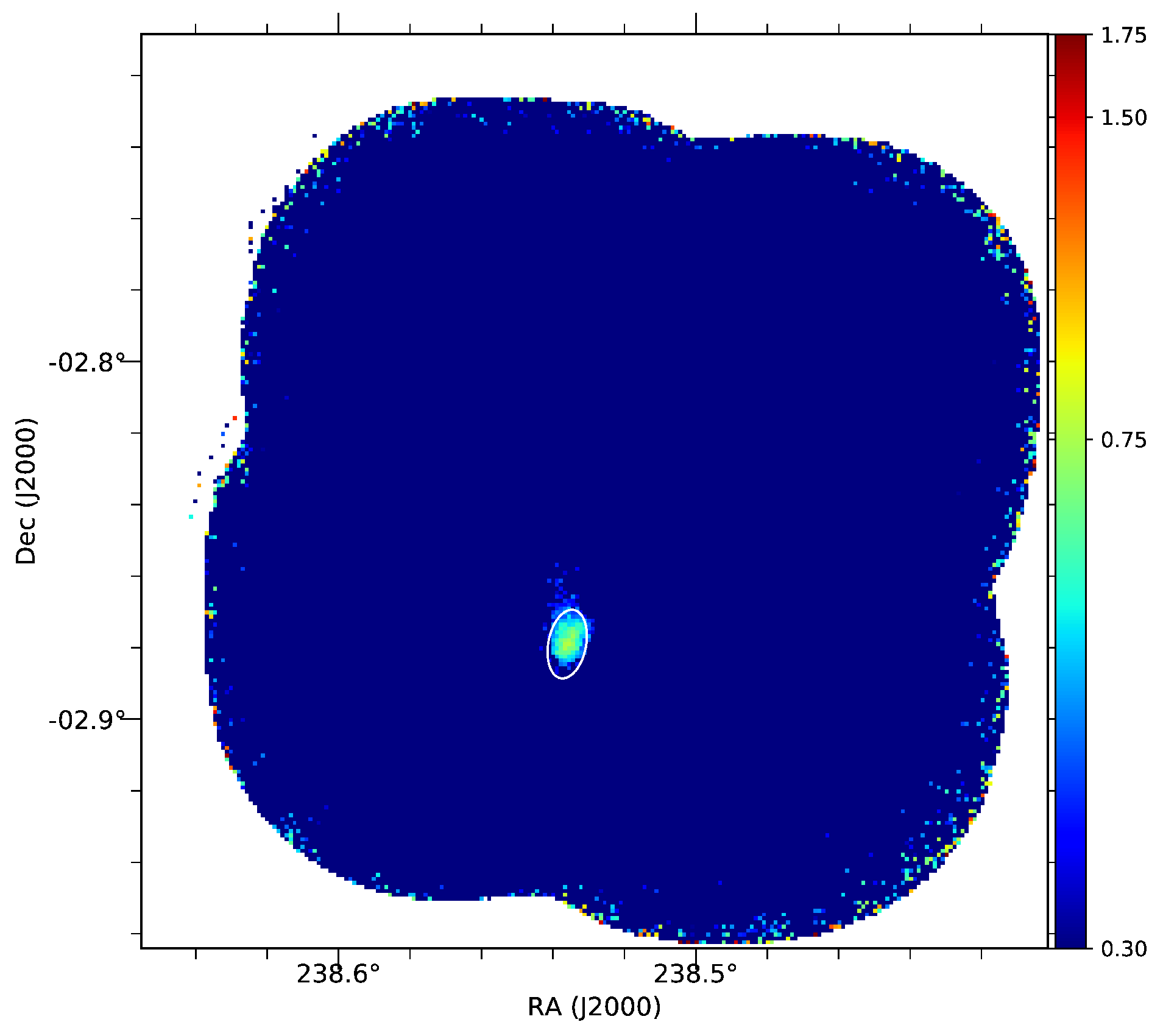}\\
\end{tabular}
\caption{Examples of observed PGCCs in the SCOPE survey. Left panel: A complicated, filamentary source at Galactic coordinates $\ell$\,=\,36$\fdg$62, $\emph{b}$\,=\,-0$\fdg$11, PGCC\_G36.62-0.11. Right panel: A high-latitude PGCC, positioned at $\ell$\,=\,6$\fdg$04, $\emph{b}$\,=\,36$\fdg$77, PGCC\_G6.04+36.77 (Liu et al., in preparation). The intensity scale in each image is mJy arcsec$^{-2}$ and the white ellipses represent the elliptical fits to the {\sc FellWalker} extractions within the observed field.}
\label{examples}
\end{figure*}

Compact sources were initially identified in 821 of the 1062 observed fields. Sources which had a peak SNR $<$ 5 were rejected, as well as sources with an aspect ratio $>$ 5. The sensitivity to extended, filamentary objects in the SCOPE survey will be explored in a future study (Fich et al., in preparation). These quality control cuts resulted in 3528 sources, with examples of the source extraction results for two fields shown in Fig.~\ref{examples}.

Table~\ref{sourcecatalogue} contains a portion of the full SCOPE compact source catalogue. The columns are as follows: (1) SCOPE catalogue source name; (2) SCOPE Region; (3) and (4) Right Ascension and Declination (J2000) of the peak flux position within the SCOPE source; (5) and (6) Right Ascension and Declination (J2000) of the central point; (7--9) semi-major axis, semi-minor axis, and position angle, measured anticlockwise from Equatorial north, of the ellipse fit to the shape of the SCOPE source which are not deconvolved sizes; (10) effective radius of source, calculated by $\sqrt{(A/\pi)}$, where $A$ is the area of the source above the detection threshold of 3$\upsigma_{\rmn{rms}}$; (11--12) peak flux density, in units of Jy\,beam$^{-1}$, and associated uncertainty; (13--14) integrated 850-$\upmu$m flux, in units of Jy, and associated uncertainty and (15) signal-to-noise ratio (SNR) of the source, calculated from the peak flux density and the $\upsigma_{\rmn{rms}}$ from the observed field. The uncertainties take account for errors in calibration, taken to be 5 per cent \citep{Dempsey13}, and uncertainties in the FCF value used, also taken to be 5 per cent. A full version is included in the Supporting Information. The 3528 sources were distributed across 558 PGCCs.

The FW routine was tested extensively within the JPS survey, with a 95 per cent recovery fraction of artificial sources found to be at approximately 5$\upsigma$. A full explanation can be found in \citet{Eden17}. Therefore, we are assuming that this completeness limit is valid for the observed SCOPE sources within the Galactic Plane. An improved recovery fraction of 99 per cent was found in two SCOPE fields at $\mid$\emph{b}$\mid$\,$>$\,30$\degr$.

\subsection{Recovered flux densities}

The FW algorithm is well understood \citep{Eden17}. Therefore, any differences in the recovered flux densities compared to other surveys are likely due to calibration issues. To test the recovered fluxes, the SCOPE fluxes were compared to those of the JPS survey \citep{Eden17}, both positionally matched and the survey as a whole.

\begin{figure*}
\begin{tabular}{ll}
\includegraphics[width=0.49\linewidth]{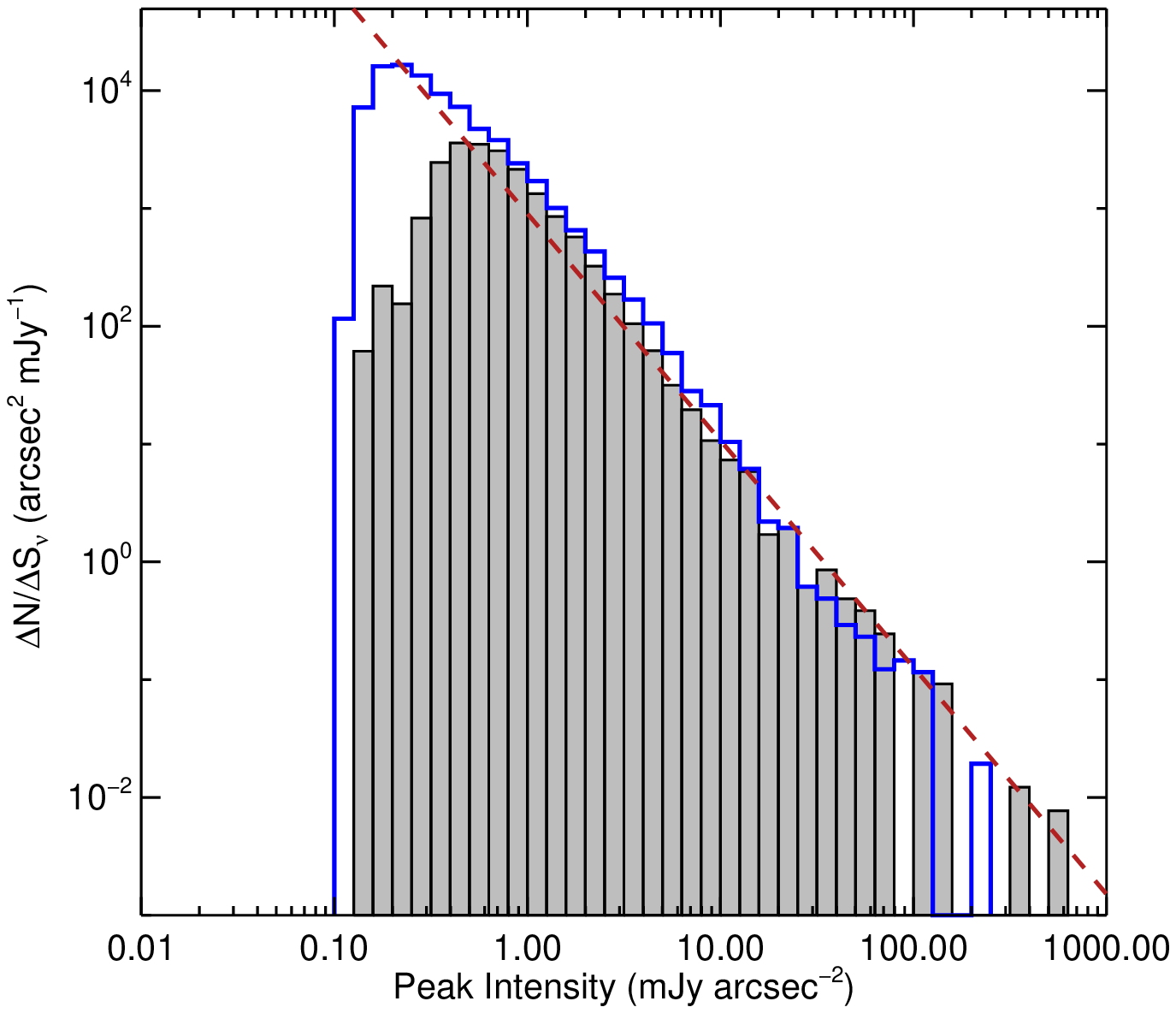} & \includegraphics[width=0.49\linewidth]{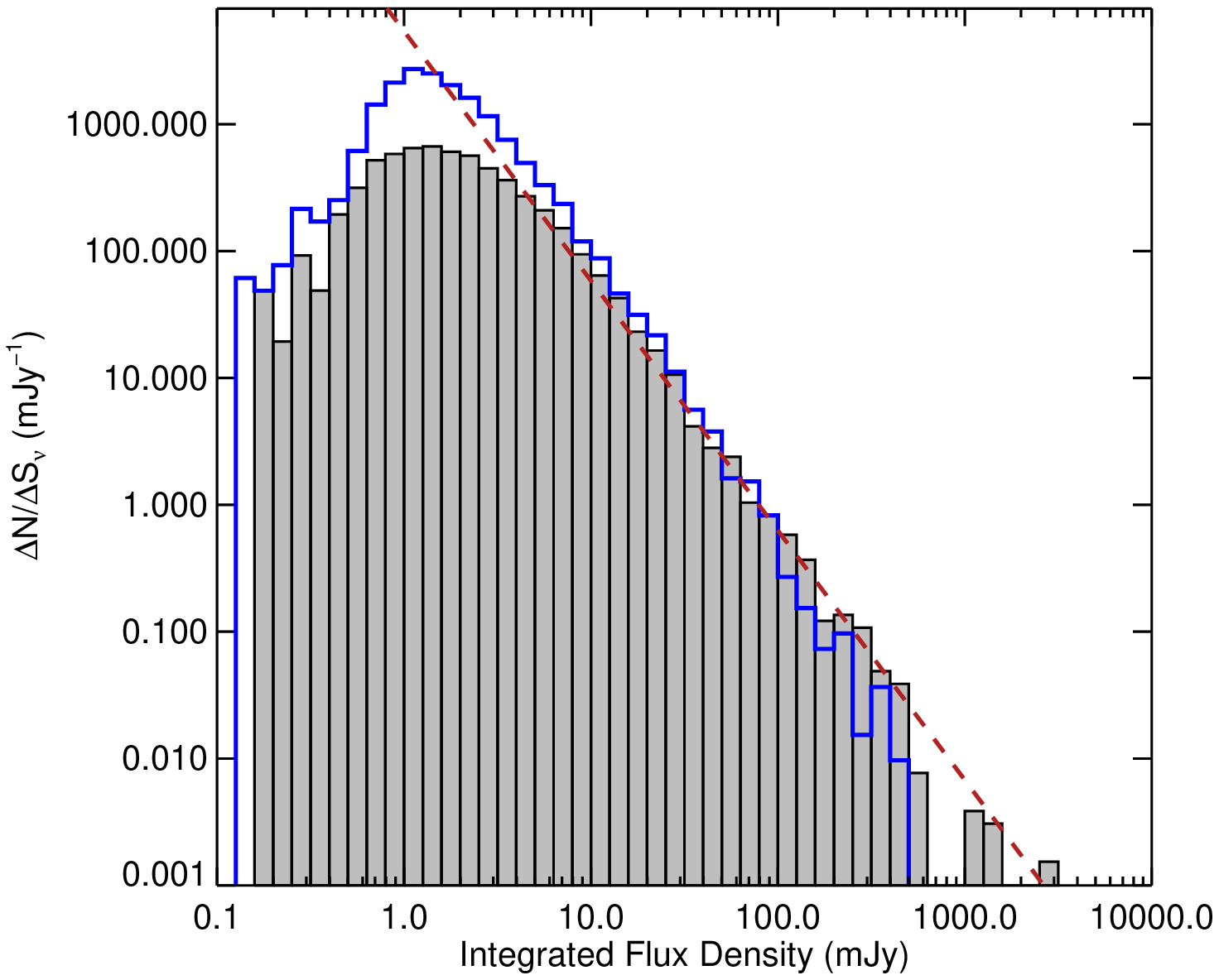}\\
\end{tabular}
\caption{Peak and integrated flux distributions for the SCOPE survey (grey, filled histogram) compared to the JPS (blue histogram) in the left and right panels, respectively. The least-squares fit to the SCOPE distributions are indicated by the red, dashed line.}
\label{fluxdist}
\end{figure*}

The distributions of peak intensites and integrated fluxes for both surveys are displayed in Fig.~\ref{fluxdist}. The JPS survey source intensities have been converted into mJy arcsec$^{-2}$ and mJy in the peak intensity and integrated flux distributions, respectively. The peak intensity distribution of the JPS goes $\sim$ 2 $\times$ deeper than SCOPE, which corresponds to the rms values of the respective surveys. The CV Daisy mode actually produces deeper observations in the central regions, with SCOPE having greater sensitivity in those regions. The peaks of the integrated flux distributions, however, are consistent with each other. By assuming single power-law laws for the tails of the distributions of the form $\Delta$$\emph{N/}$$\Delta$$S_{\nu}$\,$\propto$\,$S^{-\alpha}$, values of $\alpha$ for the two distributions were found to be $\alpha$\,=\,2.10\,$\pm$\,0.13 and $\alpha$\,=\,1.97\,$\pm$\,0.10 for the peak intensity and integrated flux distributions, respectively, above limits of 0.5\,mJy\,arcsec$^{2}$ and 2\,mJy for the peak intensity and integrated flux distributions, respectively. The SCOPE peak intensity distribution is consistent with that of the JPS ($\alpha$\,=\,2.24\,$\pm$\,0.12) but the SCOPE integrated flux distribution is flatter than that of the JPS survey ($\alpha$\,=\,2.56\,$\pm$\,0.18).

\begin{figure*}
\begin{tabular}{ll}
\includegraphics[width=0.49\linewidth]{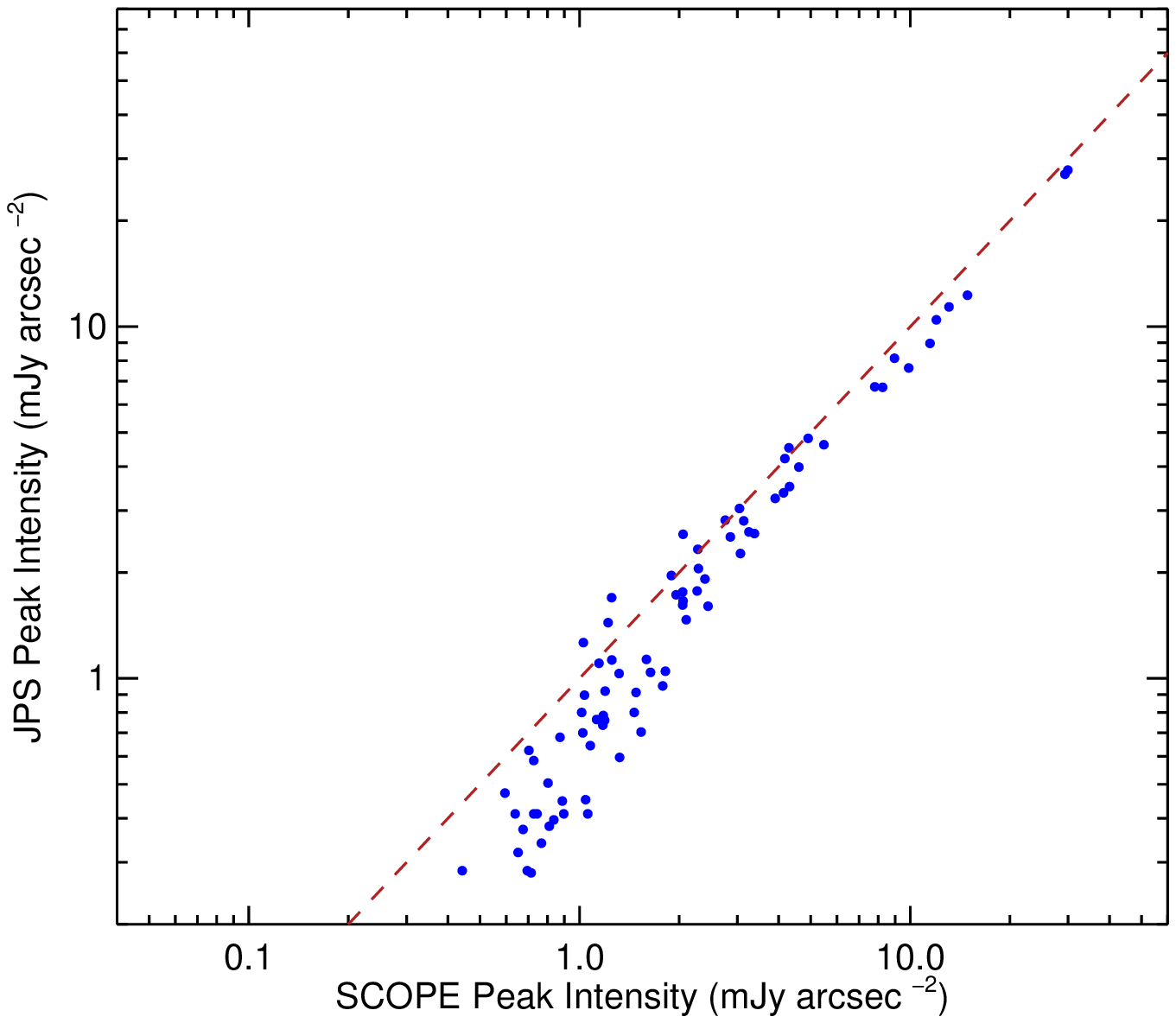} & \includegraphics[width=0.49\linewidth]{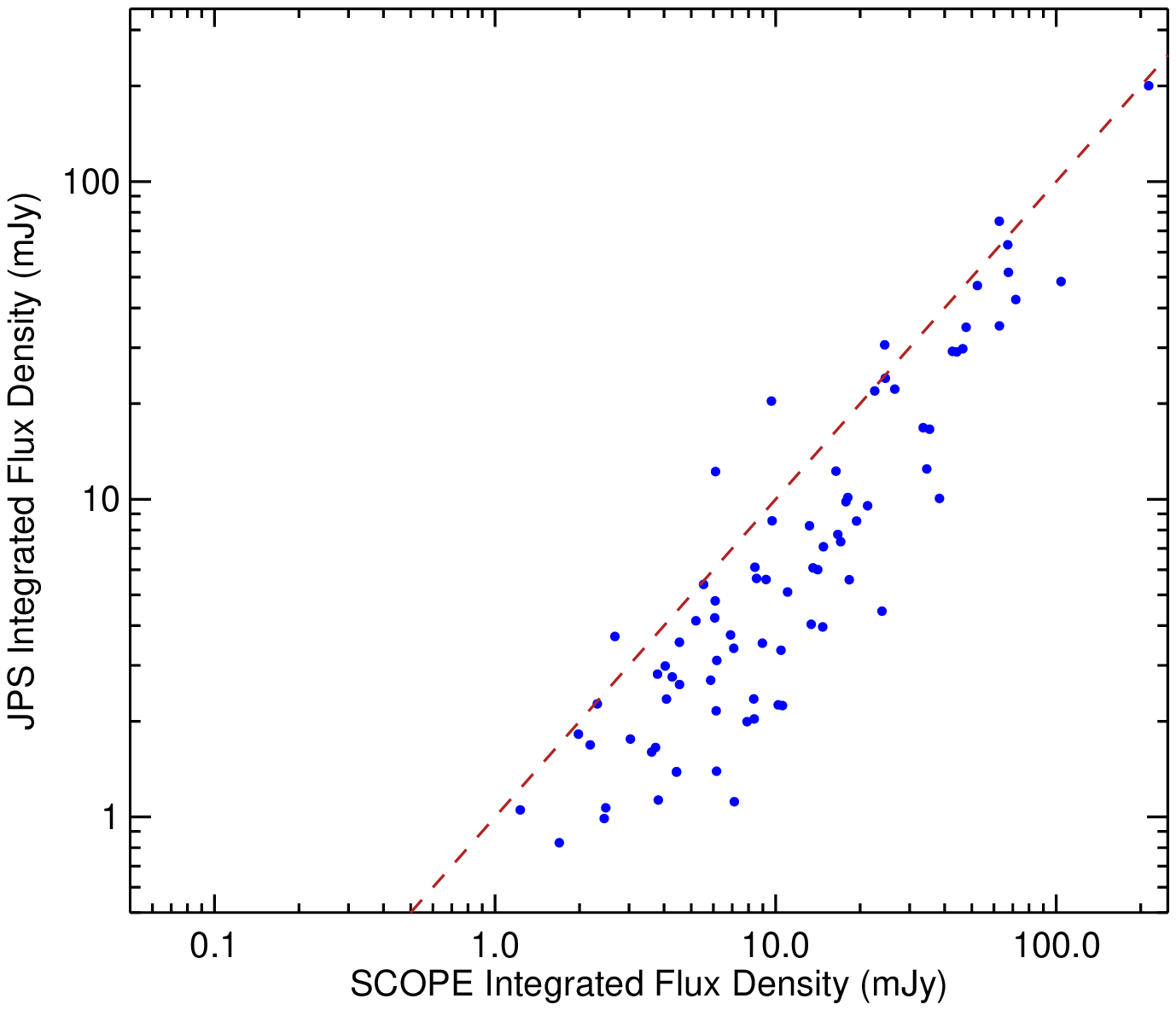}\\
\end{tabular}
\caption{Comparison of the recovered peak and total fluxes for positionally matched SCOPE and JPS sources, in the left and right panels, respectively. The red dashed line represents the 1:1 line.}
\label{fluxcomp}
\end{figure*}

There is actually a slight overlap between the SCOPE and the JPS surveys. By positionally matching the two surveys within a JCMT beam of 14.4 arcsec, we find 83 matches. A comparison of peak intensities and integrated fluxes for sources in common are shown in Fig.~\ref{fluxcomp}. The two distributions show a slight disceprency, with the integrated fluxes departing by a greater amount. The difference between the peak fluxes can be accounted for by the difference in pixel sizes (3 arcsec in the JPS, 4 arcsec in SCOPE). As reported in \citet{Mairs15,Rumble15}, changing the pixel size, especially to smaller sizes, can change the peak value, with 3-arcsec pixels giving the most accurate peak fluxes. The difference in integrated fluxes is accounted for by larger sources in the SCOPE survey, with sources in the SCOPE survey found to have a mean size 1.32 times that of the reported JPS source size. A linear best fit to the relationship gives a gradient of 1.06\,$\pm$\,0.05. The different FCFs will also account for some of the difference. However, the major difference comes from the sensitivity in the outer edges of the SCOPE maps, where these 83 sources are found. In these regions, the JPS is $\sim$ 2 $\times$ more sensitive, thus causing the sources to be broken up in the JPS. A full explanation of this effect is contained in \citet{Eden17} in the context of JPS and ATLASGAL \citep{Schuller09} comparisons.

\subsection{Angular size distribution}

\begin{figure}
\includegraphics[scale=0.5]{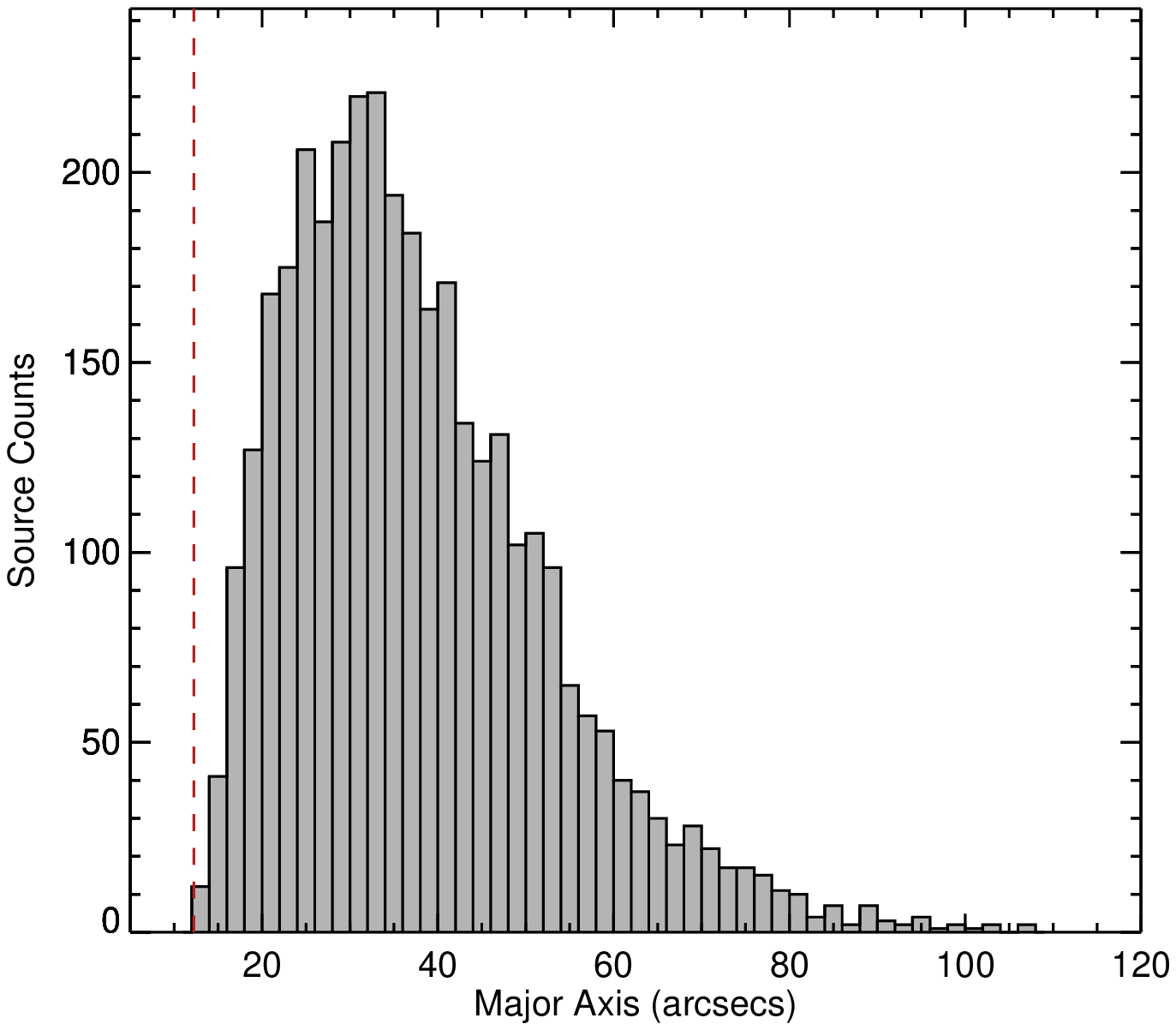}
\caption{Histogram of the major axes identified by {\sc FellWalker} of the extracted SCOPE compact sources. The red dashed line indicates the beam size of the JCMT at 850\,$\upmu$m.}
\label{angsize}
\end{figure}

\begin{figure}
\includegraphics[scale=0.5]{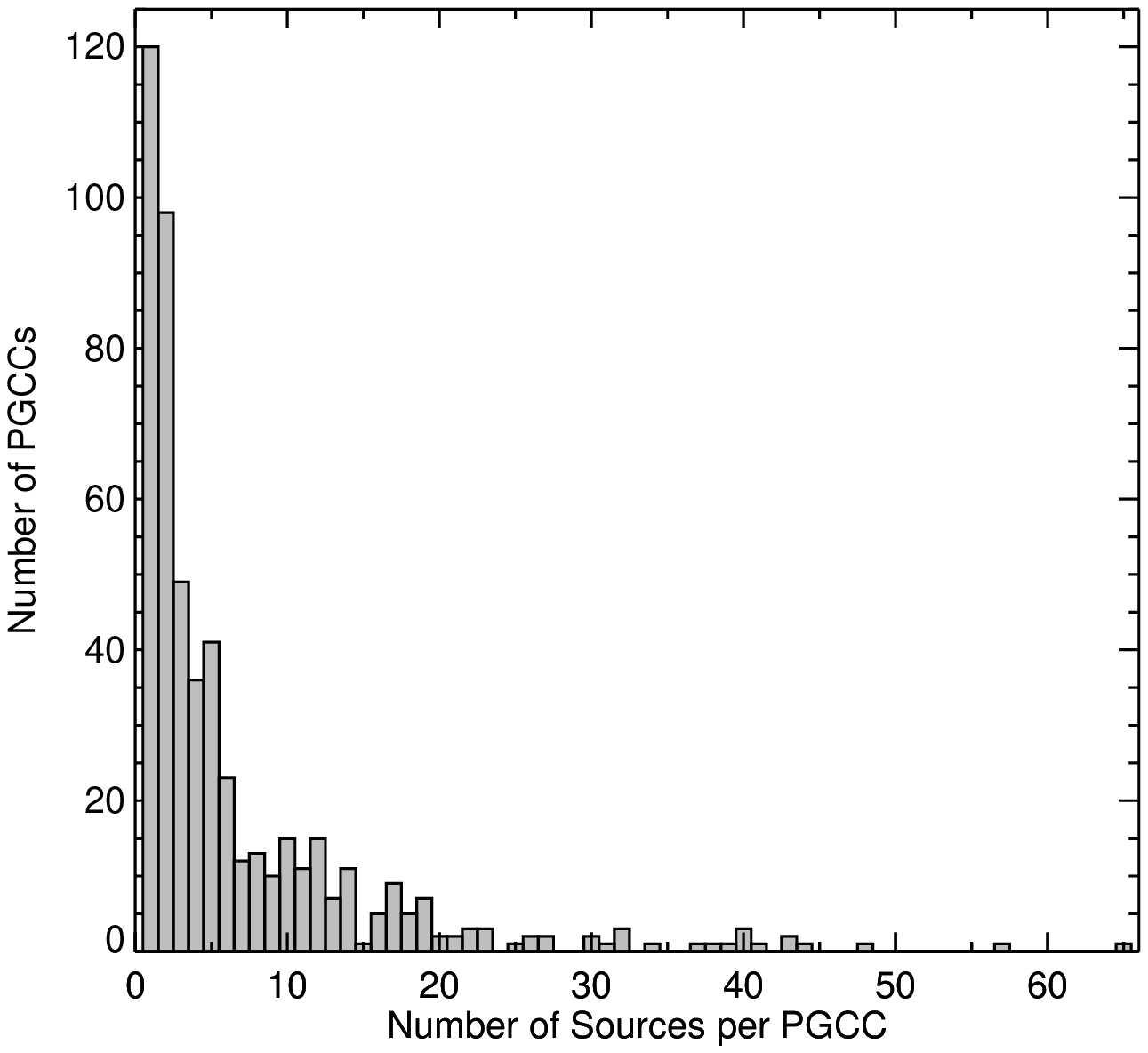}
\caption{Histogram of the number of compact sources extracted per PGCC with a detection in the SCOPE survey.}
\label{multiplicity}
\end{figure}

The angular size distribution of the SCOPE sources is shown in Fig.~\ref{angsize}. The plotted quantity is the major axis of the elliptical fit to the source, provided by FW. The reported sizes are not deconvoled sizes. The peak of the distribution, found at 35 arcsec, is in marked contrast to the peak at 8 arcmin found in the \emph{Planck} catalogue \citep{Planck16}. This difference further exemplifies the inner substructure identified by the higher resolution SCOPE survey and the presence of multiple SCUBA-2 sources inside a single PGCC, which is highlighted in Fig.~\ref{multiplicity}, where 61 per cent of detected PGCCs have 3 or more SCOPE sources.

\begin{figure}
\includegraphics[scale=0.5]{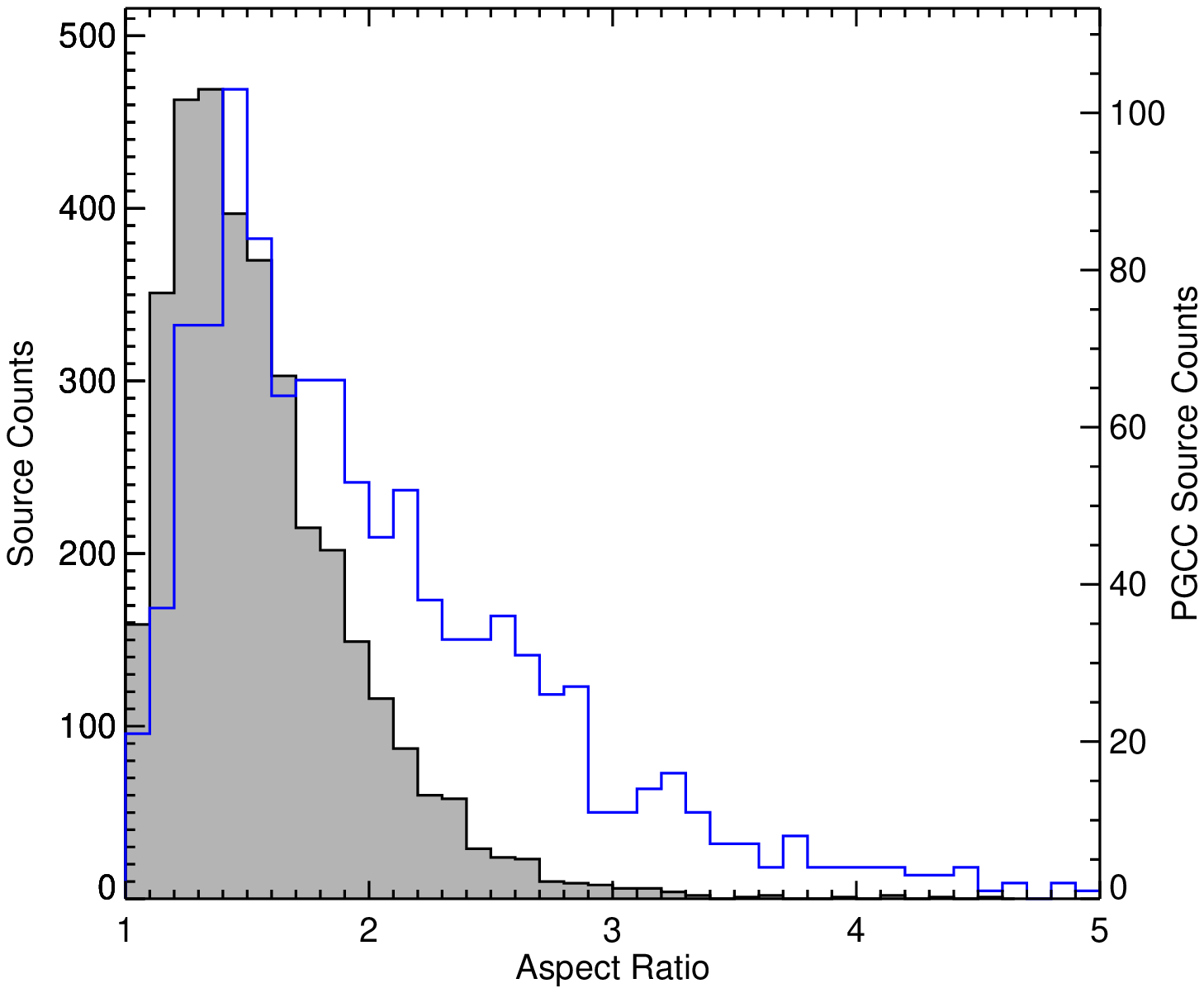}
\caption{Histogram of the aspect ratios of all SCOPE sources (grey histogram) with the aspect ratios of the SCOPE-observed PGCCs overlaid in the blue histogram.}
\label{aspect}
\end{figure}

We present the aspect ratios of the SCOPE sources in Fig.~\ref{aspect}. We have overlaid the aspect ratios of the observed \emph{Planck} PGCCs in this survey. \citet{Planck16} found that 40 per cent of sources had an aspect ratio of between 2 and 3, with a shoulder at those values not present in the SCOPE compact source catalogue. This shoulder in the PGCC catalogue aspect ratios is a hint at the substructure observed by SCOPE, with a higher aspect ratio pointing towards more filamentary structures. It is also a reflection of the nature of the sources extracted by the FW algorithm, which is attuned more towards extracting compact objects.

\section{SCOPE Survey and Data Access}

The SCOPE data products can be downloaded from the Canadian Astronomy Data Centre's JCMT Science Archive\footnote{http://www.cadc-ccda.hia-iha.nrc-cnrc.gc.ca/en/jcmt/}, with the proposal IDs MJLSY14B, M15AI05, M15BI06, and M16AL003. These IDs correspond to observations taken in the SASSy survey, two PI proposals, all of which formed the pilot observations, and the SCOPE survey, respectively. As well as these data, the raw observation data can also be downloaded from the same location.

The full source compact source catalogue is available as Supporting Information.

\section{Results}

\subsection{Detection statistics}

\begin{figure*}
\begin{tabular}{lll}
\includegraphics[width=0.32\linewidth]{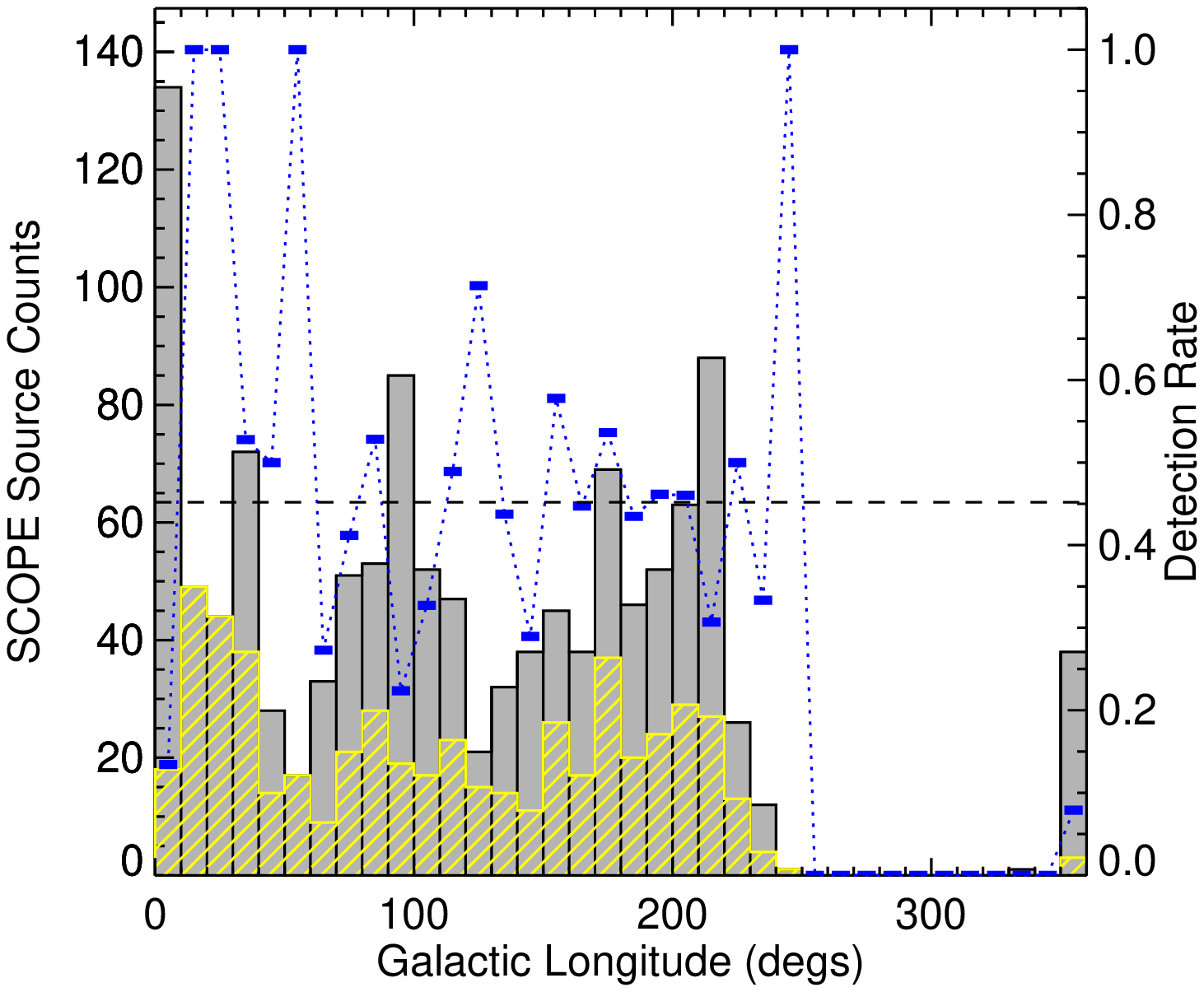} & \includegraphics[width=0.32\linewidth]{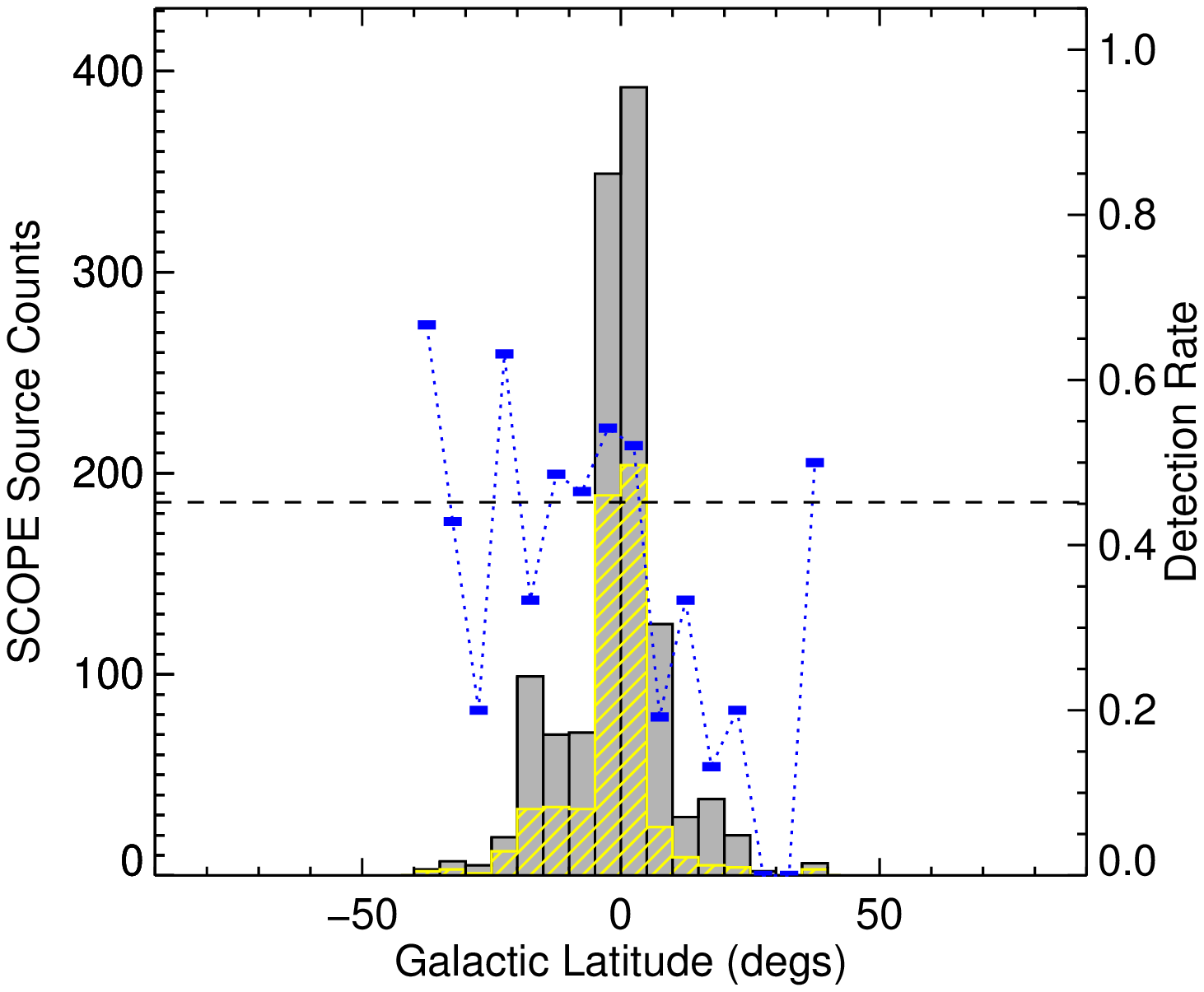} & \includegraphics[width=0.32\linewidth]{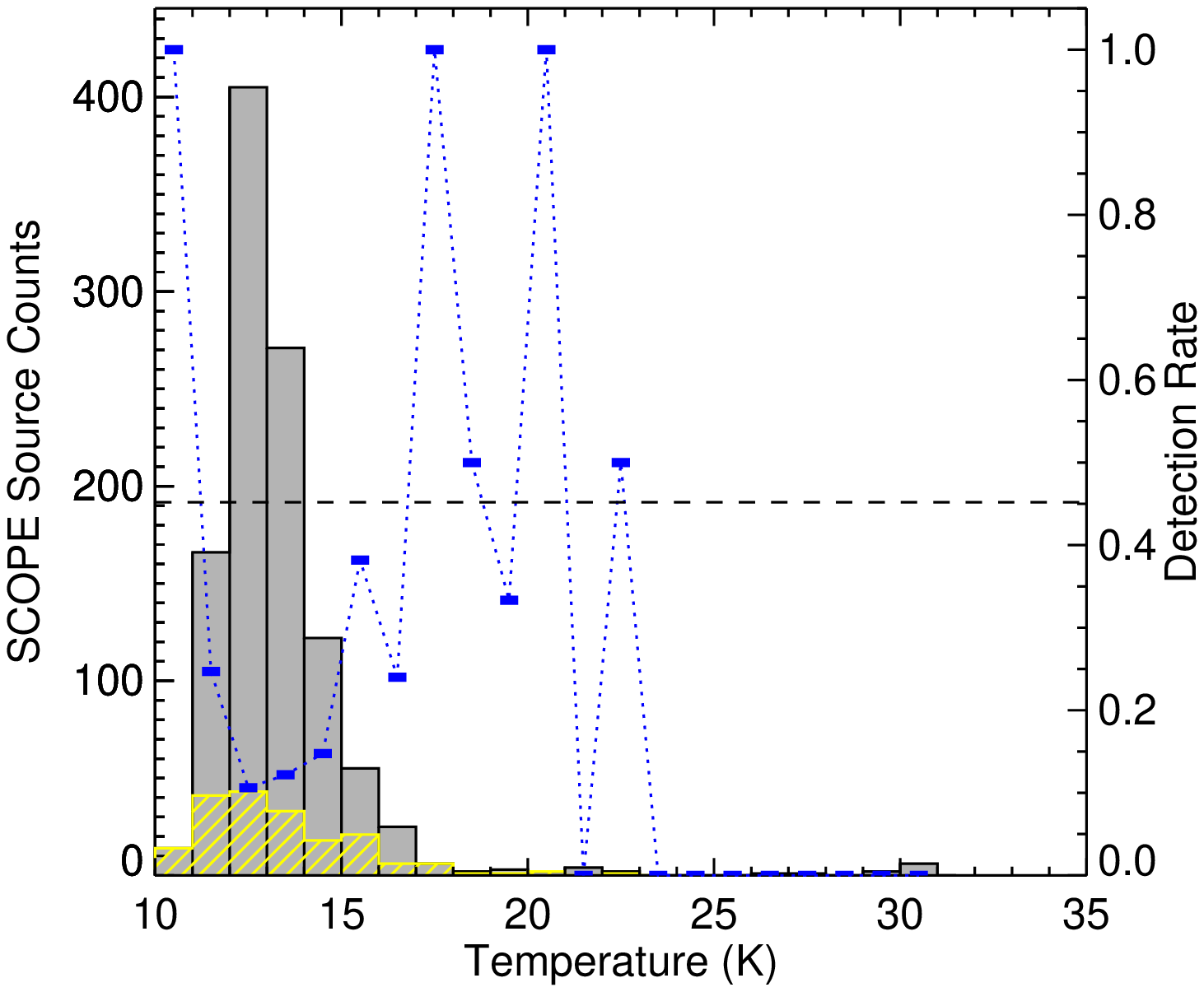}\\
\includegraphics[width=0.32\linewidth]{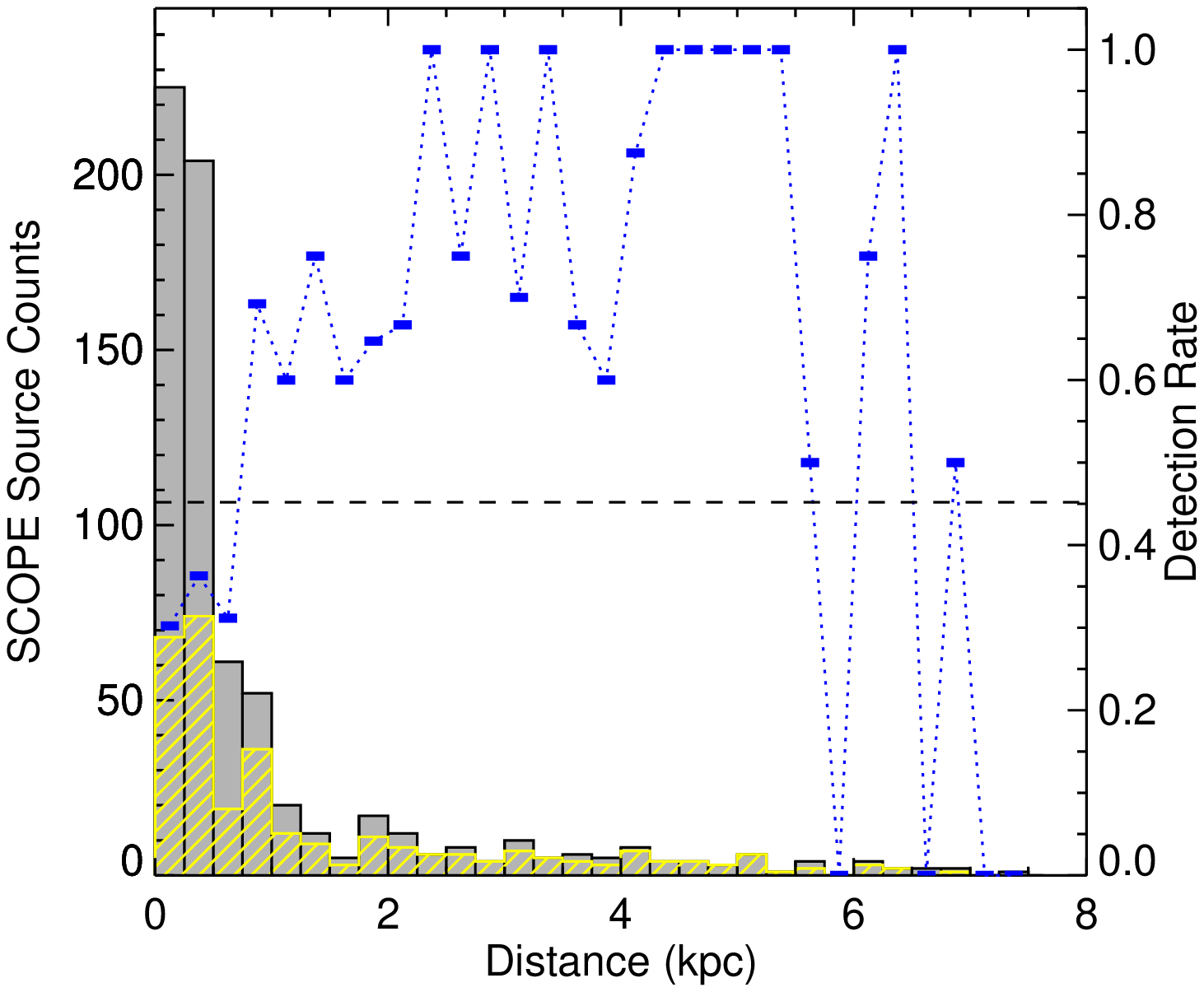} & \includegraphics[width=0.32\linewidth]{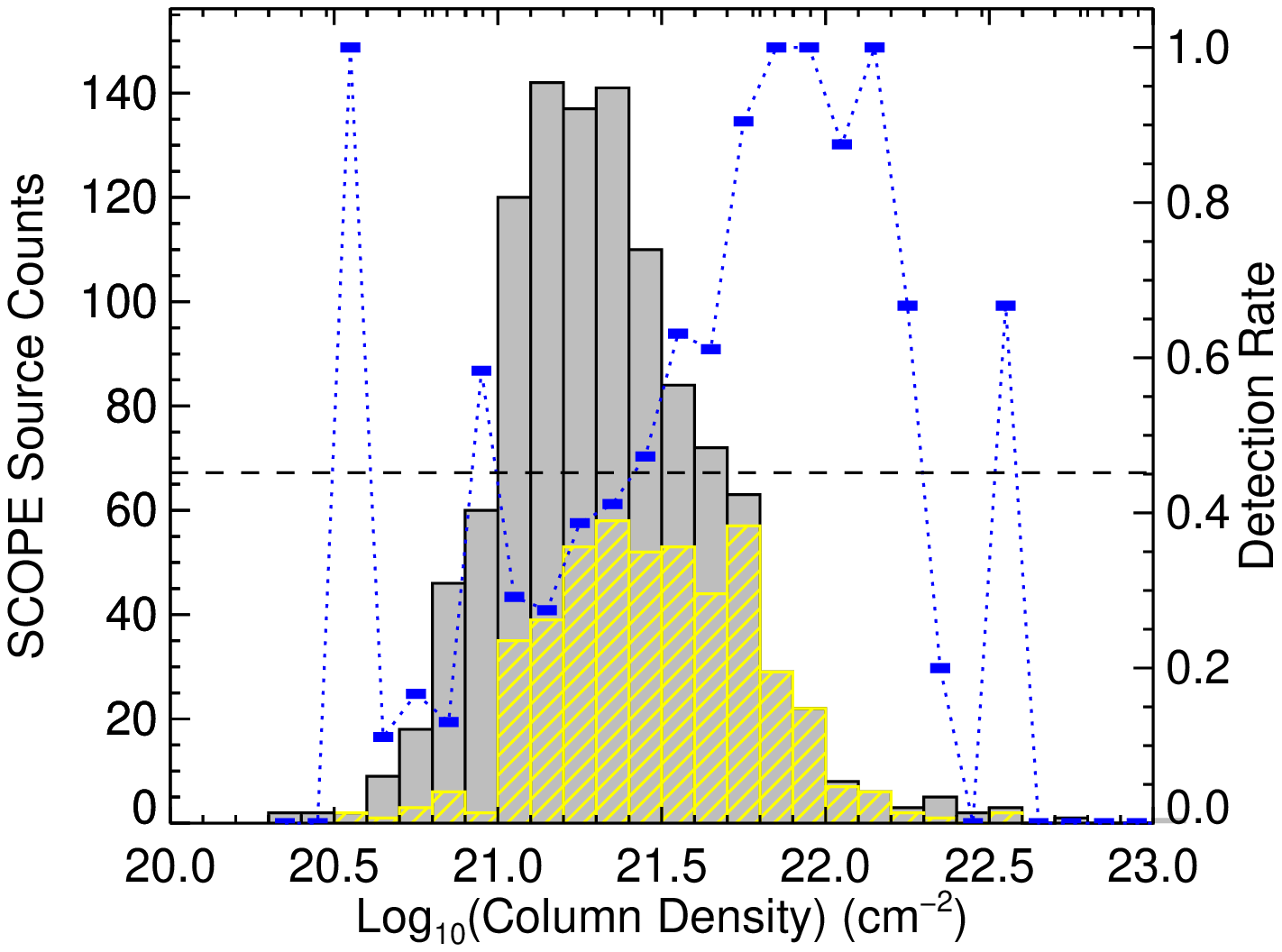} & \includegraphics[width=0.32\linewidth]{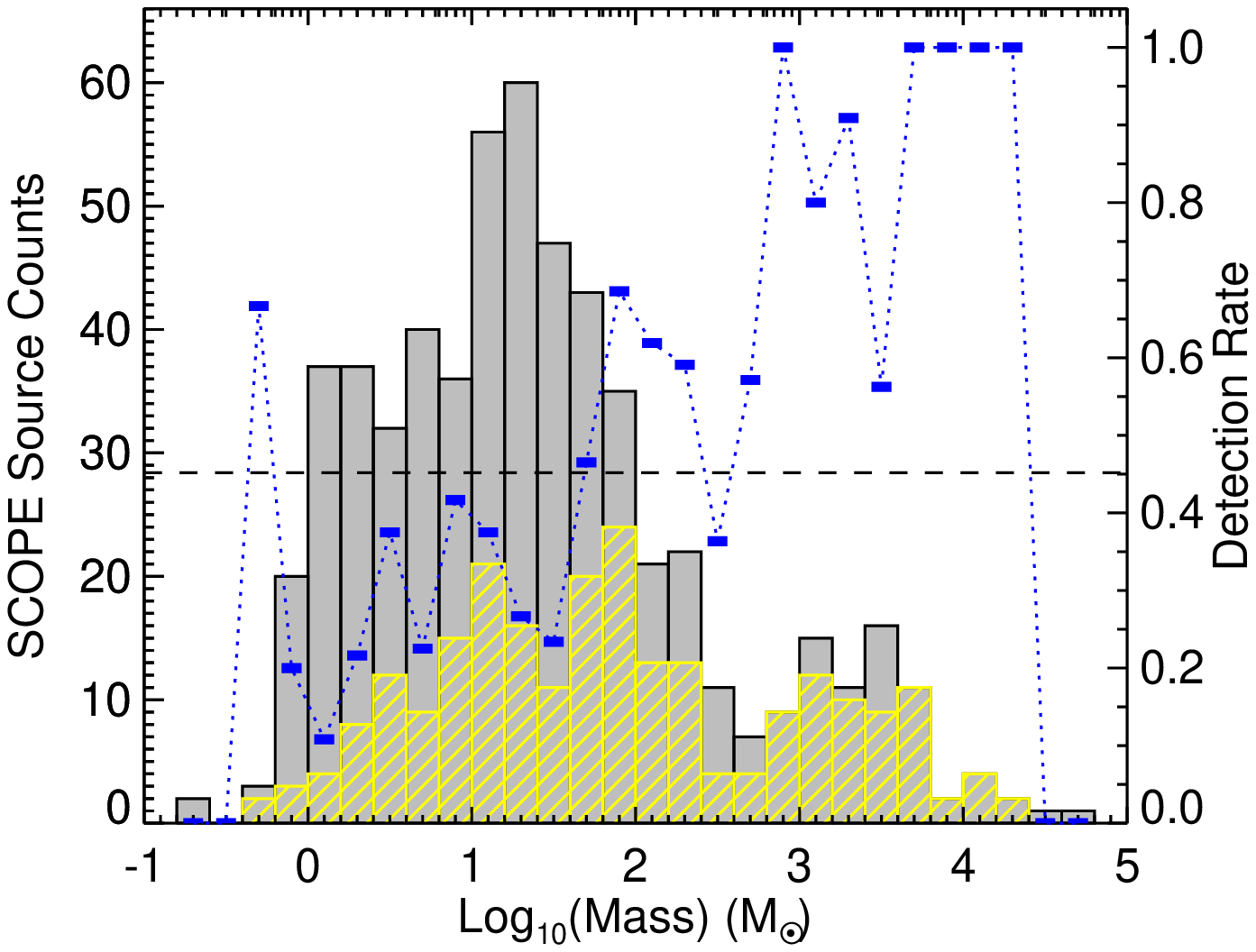}\\
\includegraphics[width=0.32\linewidth]{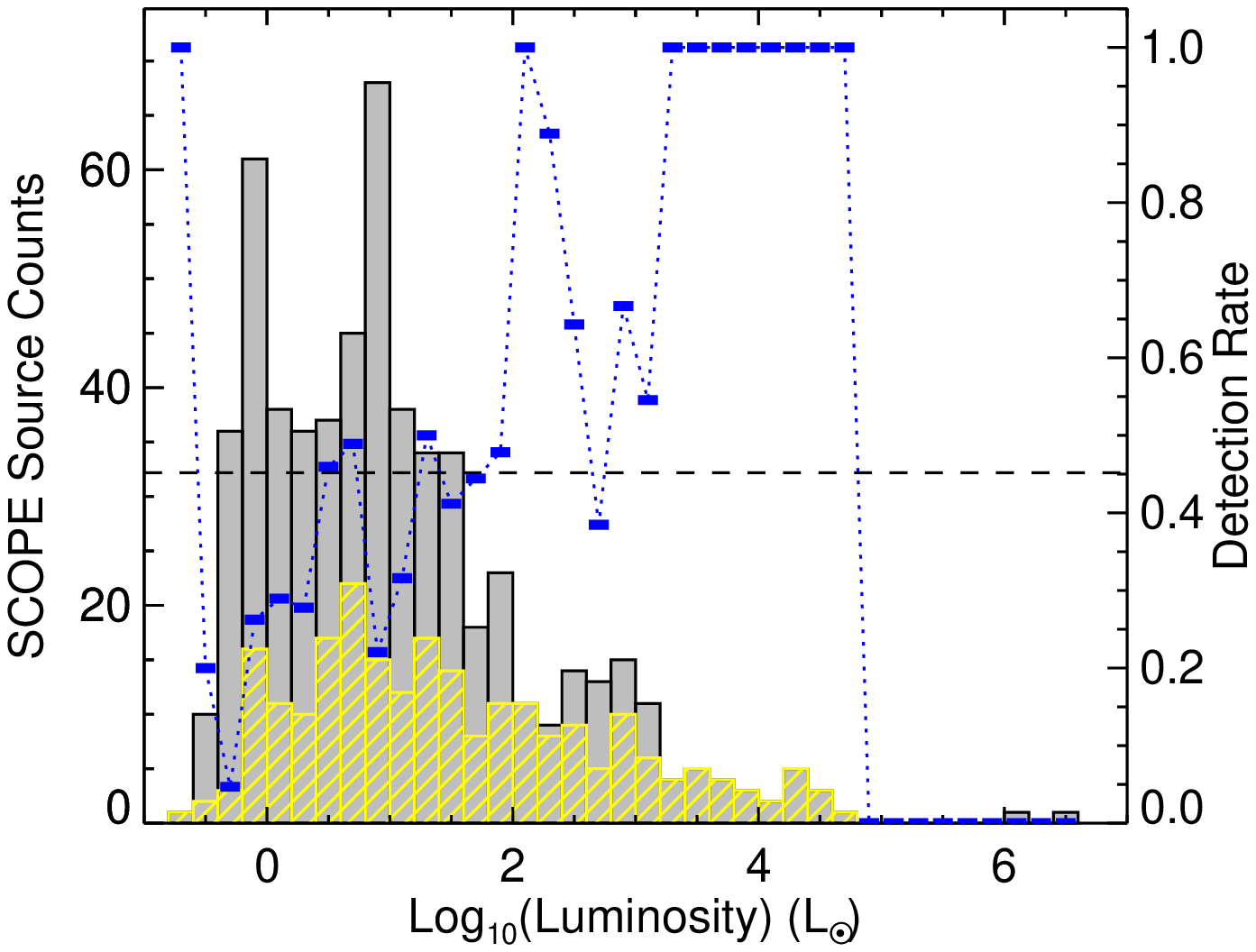} & \includegraphics[width=0.32\linewidth]{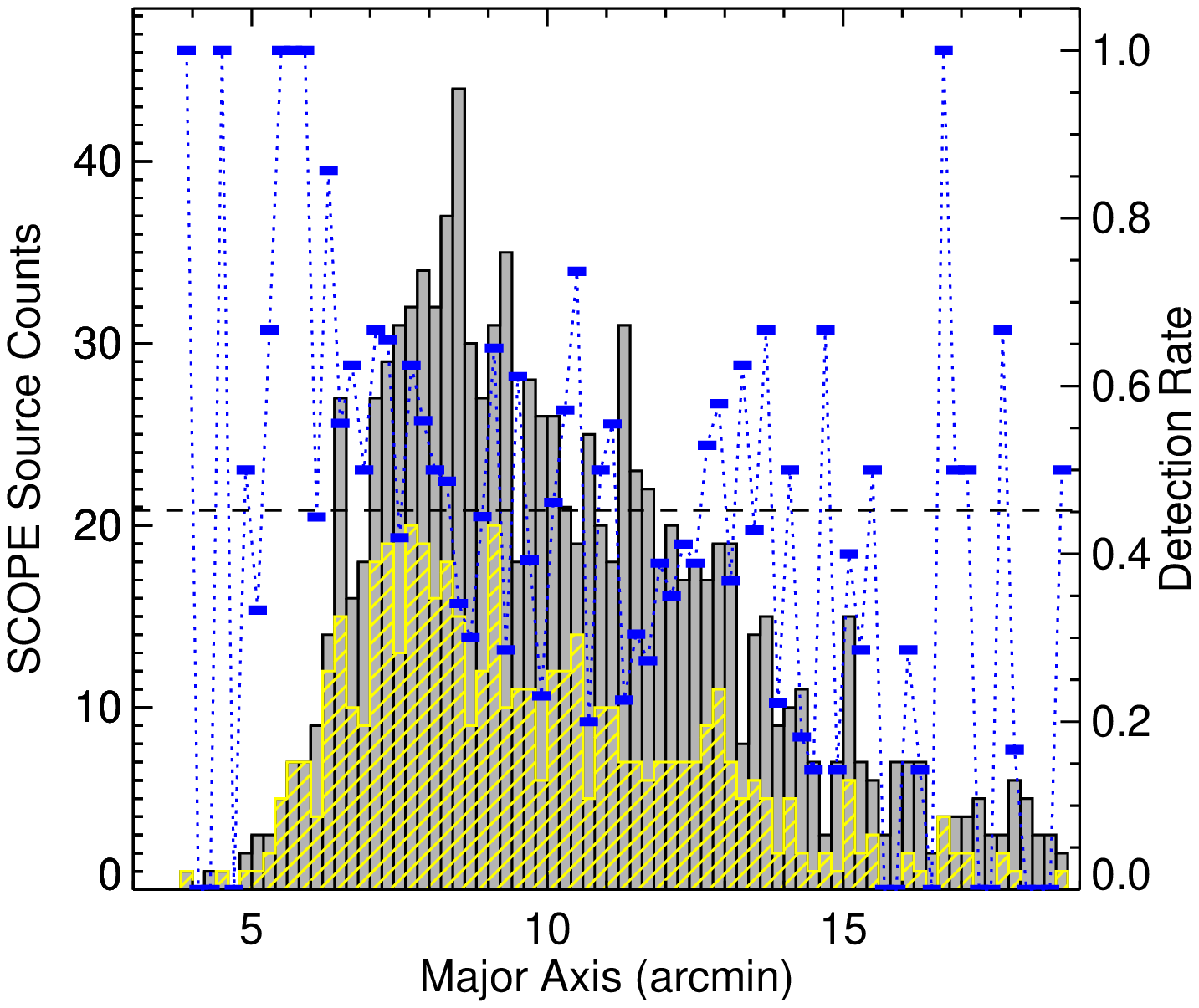} & \includegraphics[width=0.32\linewidth]{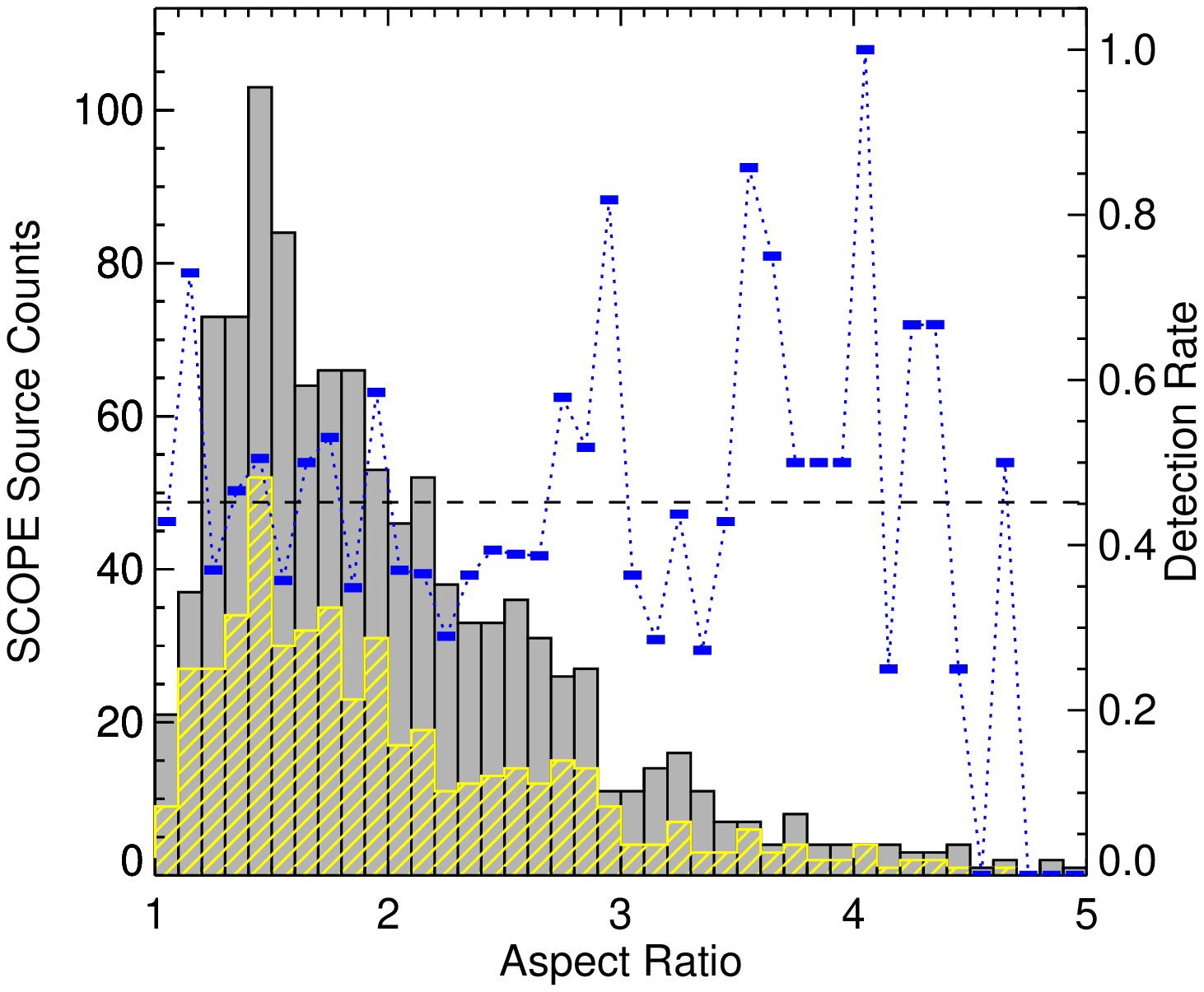}\\
\end{tabular}
\caption{Distributions of the observed, $\emph{Planck}$-derived statistics of SCOPE sources (grey histograms) overlaid with the detected PGCCs (yellow hashed histogram). The ratio of the two histograms is overlaid by the blue dotted line. The black dashed line respresents the overall detection rate. As in Fig.~\ref{statistics}, the top row displays the Galactic longitude, Galactic latitude, and temperature, in the left, central, and right panels, respectively. The middle row are the distance, column density, and mass, with the bottom row showing the luminosity, major axis, and aspect ratio.}
\label{detectionratios}
\end{figure*}

The PGCCs observed in the SCOPE survey were chosen to sample across the spectrum of a host of statistics, as displayed in Fig.\ref{statistics}, with a bias towards the highest column density sources, as described earlier. The detection rate of each statistic is shown in Fig.\ref{detectionratios}. Six high-latitude PGCCs with detections ($\mid$\emph{b}$\mid$\,$>$\,30$\degr$) were excluded from the statistics as they were found to be associated with lensed galaxies (Liu et al., in preparation).

The longitude detection rates are approximately equal to the overall detection rate (0.46), with lower rates found in the central 20 degrees ($\mid$\emph{b}$\mid$\,$<$\,2$\degr$). In these longitudes, higher latitude sources are observed, with lower detection rates found outside of the central 4 degrees ($\mid$\emph{b}$\mid$\,$>$\,2$\degr$) of the Galactic Plane. In the latitude range of $\mid$\emph{b}$\mid$\,$<$\,2$\degr$, the detection rate is slightly higher than the overall rate, approximately 0.50. This is also reflected in the detection rate as a function of distance. The detection rate at higher distances, those classically taken to be in the Galactic Plane, are well detected. However, the local, higher latitude sources are not detected at as higher rate.

The column densities are 95 per cent complete above densities of $N_{\rmn{H_{2}}}$\,$>$\,5\,$\times$\,10$^{21}$\,cm$^{-2}$. This detection rate corresponds to the threshold for star formation found in studies of nearby star-forming clouds \citep{Andre10,Heiderman10,Lada10,Andre14} Some bins are subject to low detection rates due to low number statistics. The mass and lumniosities are complete to a 95 per cent rate above 5\,$\times$\,10$^{3}$\,M$_{\odot}$ and 1\,$\times$\,10$^{3}$\,L$_{\odot}$, respectively.

\begin{figure*}
\begin{tabular}{ll}
\includegraphics[width=0.49\linewidth]{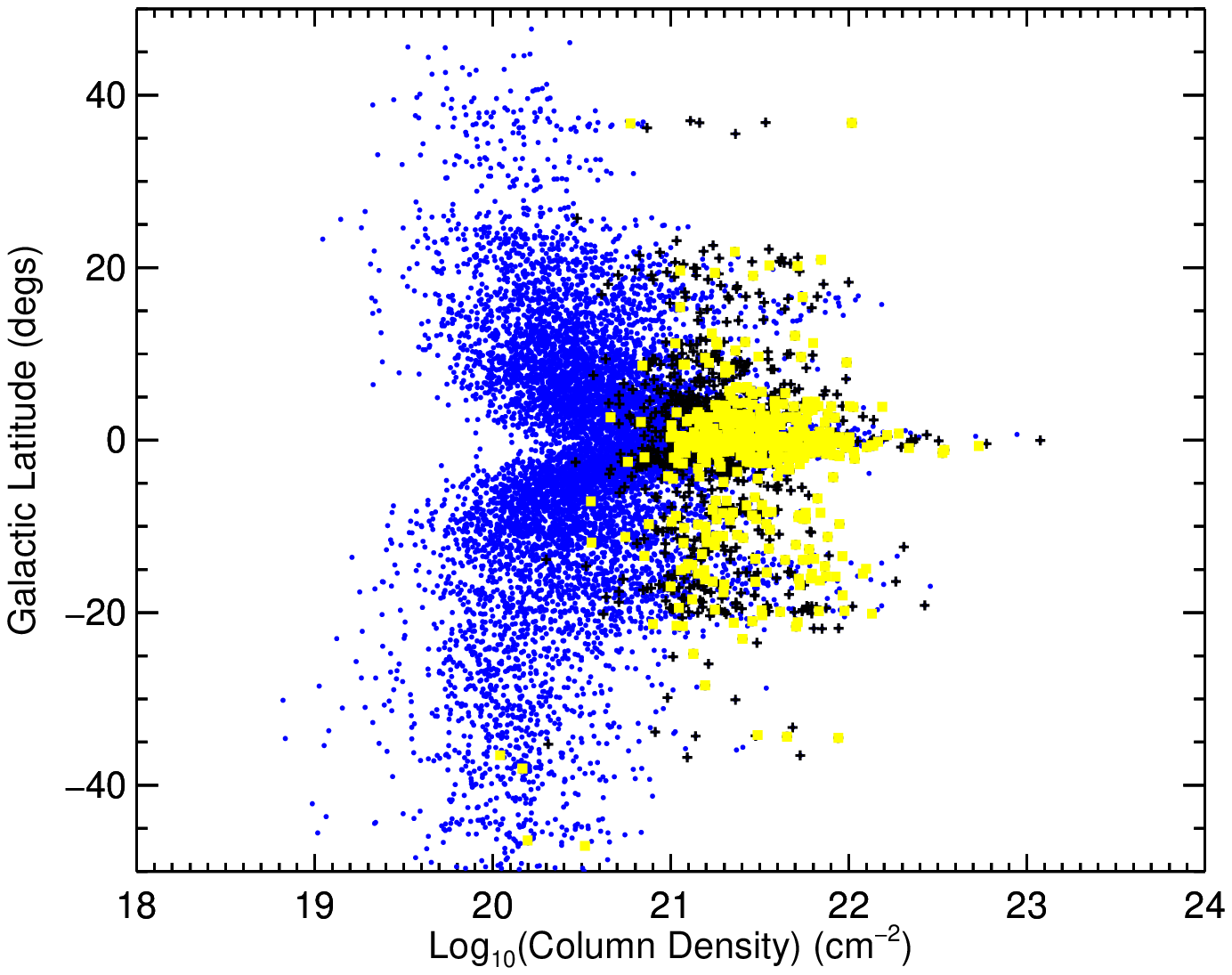} & \includegraphics[width=0.49\linewidth]{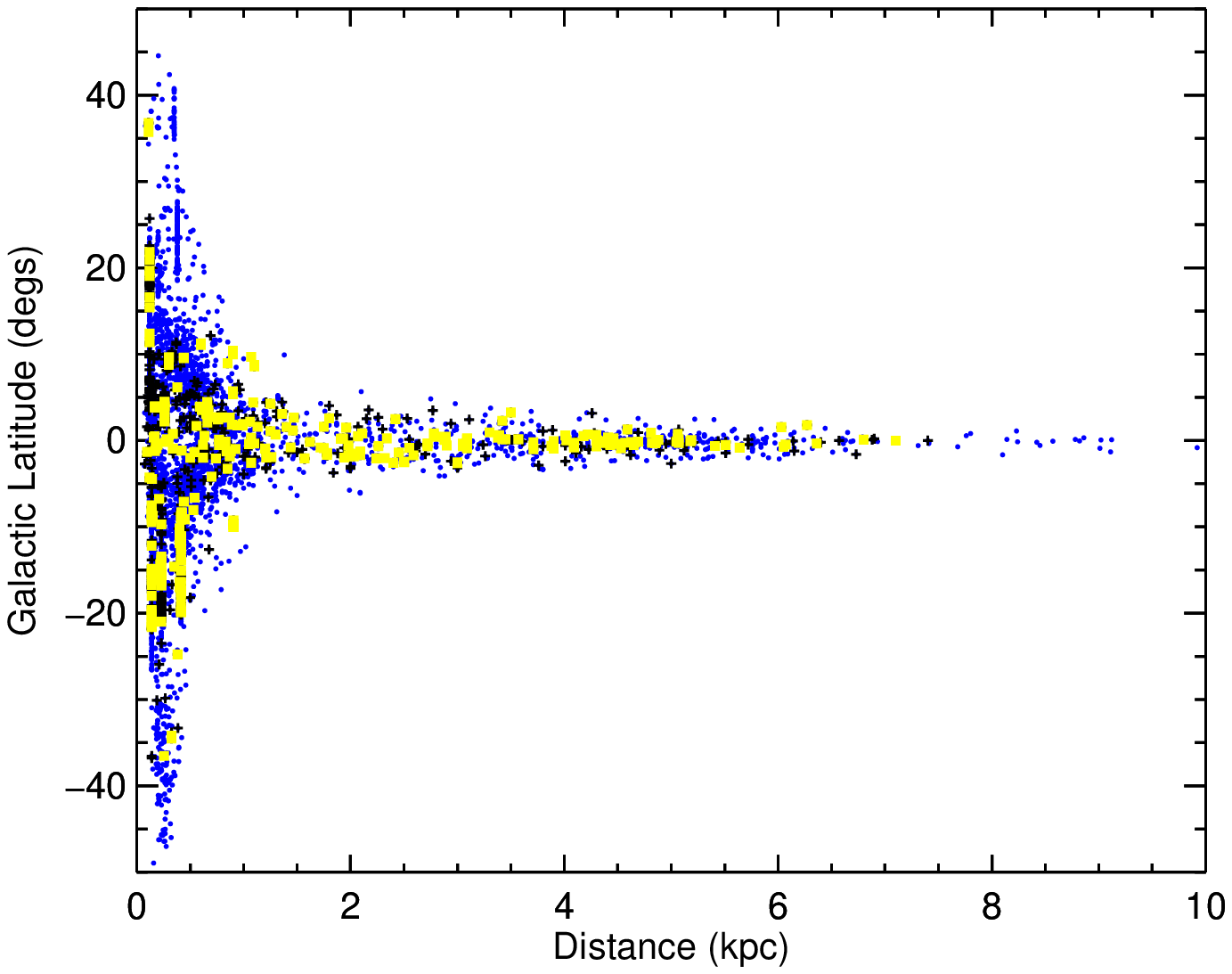}\\
\end{tabular}
\caption{Latitude of PGCCs against column density (left panel) and distance (right panel). The blue circles are the entire PGCC sample from \citet{Planck16}, the black plus symbols are the observed SCOPE PGCCs, with the yellow squares representing the detected SCOPE PGCCs.}
\label{coldenlatitude}
\end{figure*}

The column density detection rate does not account for the selection bias imposed on the initial sample. When comparing the entire PGCC population, the observed SCOPE sample, and the detected PGCCs in SCOPE, we can see that the highest column density sources are confined to the lower latitudes, which are the PGCCs with the largest distances, as seen in Fig.~\ref{coldenlatitude}. Deeper observations will be required to trace the lowest column density objects.

The major axes and aspect ratios fluctuate around the actual detection rate of the SCOPE survey.

The detection rate of the temperatures is skewed by the lack of derived temperatures by \citet{Planck16}. 619 of the 1235 observed PGCCs do not have derived temperatures, with 360 of those 619 detected in SCOPE. The detection rate at the lower temperatures of less than 15\,K is 0.15.

\subsection{Star formation out of the Galactic Plane}

The SCOPE survey gives a sample of potentially star-forming cores and clumps in different Galactic environments. The YSO catalogue of \citet{Marton16} provides an all-sky catalogue derived from the AllWISE catalogue \citep{Cutri13}. By positionally matching these two catalogues, we find 865 YSOs located within the map extents of 201 observed PGCCs that have detected compact sources.

The YSO catalogue of \citet{Marton16} contains the magnitudes of the 4 WISE bands (3.4\,$\upmu$m, 4.6\,$\upmu$m, 12\,$\upmu$m, and 22\,$\upmu$m) as well as the $J$, $H$, and $K$ bands from the positional matching of 2MASS Point Sources \citep{Cutri03}. These 7 bands can be used to calculate the luminosities of each YSO, and therefore the total YSO luminosity associated with a PGCC. The luminosities are calculated, once the magnitudes are converted into fluxes, using a trapezium rule estimation in log-log space, which was shown to provide a good approximation of the luminosity \citep{Eden15}, with other studies using this method \citep[e.g.,][]{Veneziani13}.

These luminosities, $L$, can be compared to the masses, $M$, of the SCOPE detected YSOs to determine the ratio of $L/M$, a measure of the current star formation and an indicator of the evolutionary state of that star formation \citep[e.g.,][]{Elia17,Urquhart18}. Comparing this ratio between the nearby sources, complementary to the Gould's Belt, and the more distant Galactic Plane can determine whether the star formation in either environment is at a different evolutionary stage (on average).

The masses were derived from the total emission within the SCOPE maps. The masses in the SCOPE maps are estimated using the optically thin approximation:

\begin{equation}
M = \frac{S_{\nu}D^{2}}{\kappa_{\nu}B_{\nu}(T_{d})}
\end{equation}

\noindent where $S_{\nu}$ is the integrated flux of the emission, $D$ is the distance (omitted in the calculation for the ratio of $L/M$), $\kappa_{\nu}$ is the mass absorption coefficient taken to be 0.01\, cm$^{2}$\,g$^{-1}$ \citep{Mitchell01} which accounts for a gas-to-dust ratio of 100, and $B_{\nu}(T_{d})$ is the Planck function evaluated at temperature, $T_{d}$, taken to be 13\,K, the peak of the SCOPE-observed PGCC-temperature distribution.

The sample of 201 PGCCs with a YSO was split into two populations, a Gould's Belt-like population and the distant Galactic Plane. This was done by using a distance cut of 0.5\,kpc, as this is taken to be the furthest Gould's Belt sources \citep{Ward-Thompson07}, with distances below this in the Gould's Belt, and greater distances in the Galactic Plane. From hereafter, these populations will be referred to as the nearby and distant samples, respectively. The distances were estimated in three ways. The first was to take the distances as derived by \citet{Planck16}. This accounted for 91 PGCCs. The second was to extract spectra from spectral line data from existing surveys \citep{Dame01,Jackson06,Dempsey13a,Rigby16}, with sources in the Plane compared to the Galactic rotation curve of \citet{Brand93}. This accounted for a further 108 PGCCs. The final two were well out of the Plane at high latitudes, and were assumed to be local.

\begin{figure}
\includegraphics[scale=0.5]{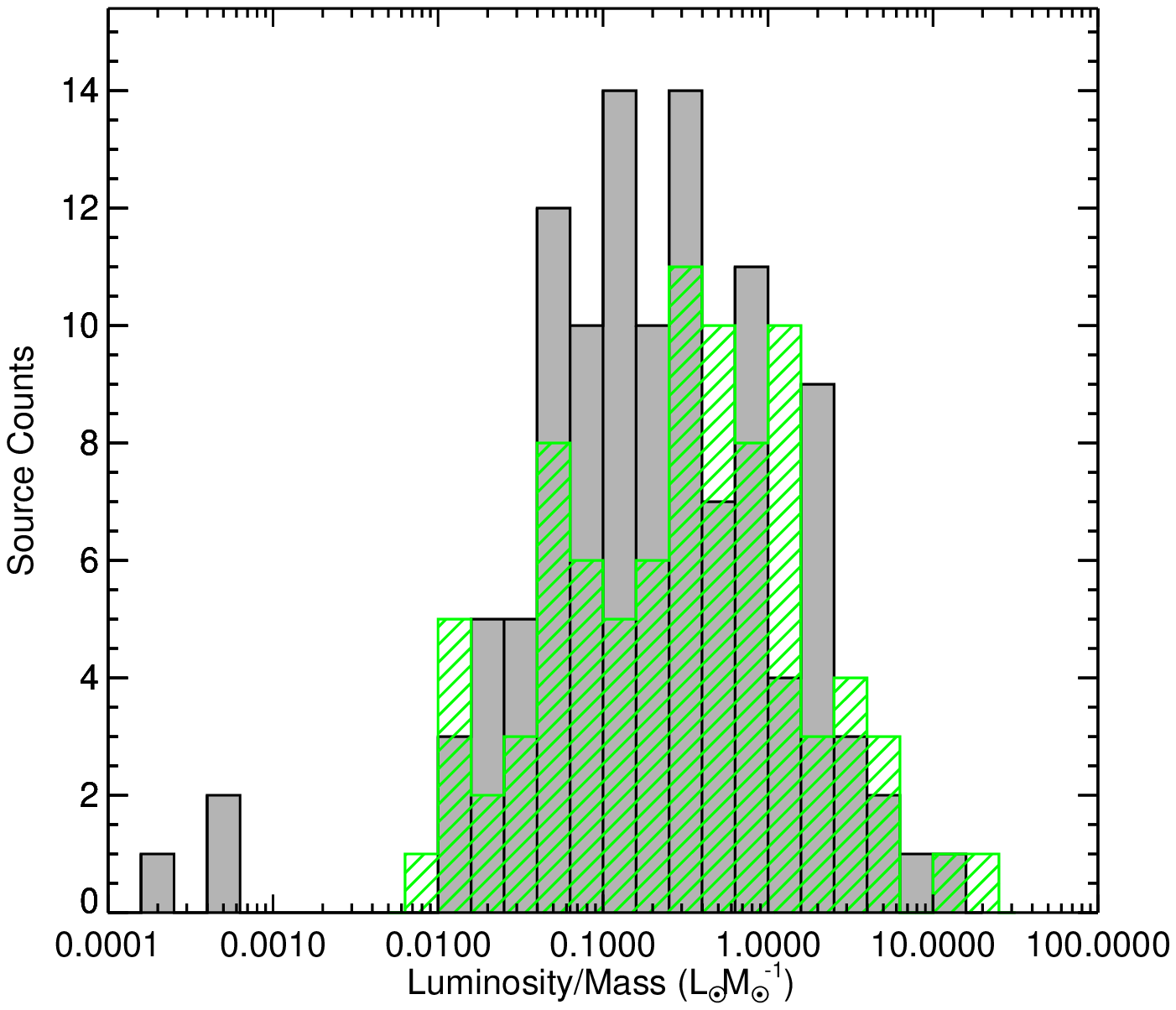}
\caption{Histogram of the $L/M$ ratios in SCOPE detected PGCCs in the further (grey) and in the nearby (green) Galactic Plane.}
\label{lm}
\end{figure}

 We found 114 and 87 PGCCs in the distant and nearby samples, respectively. The distribution of the $L/M$ ratios for each of these populations is shown in Fig.~\ref{lm}. The mean values of the two populations are 0.78\,$\pm$\,0.16\,L$_{\odot}$/M$_{\odot}$ and 1.14\,$\pm$\,0.30\,L$_{\odot}$/M$_{\odot}$ for the distant and nearby samples, respectively, with median values of 0.21\,$\pm$\,0.16\,L$_{\odot}$/M$_{\odot}$ and 0.32\,$\pm$\,0.28\,L$_{\odot}$/M$_{\odot}$, respectively. The populations are consistent with each other, within the errors, indicating that the star formation in the distant Galactic Plane is at roughly the same evolutionary stage as that within the nearby Gould's Belt-like sample. A Kolomogorov--Smirnov (K--S) test of the two subsamples finds a 25 per cent chance they are not drawn from the same distribution. We therefore cannot strongly reject the null hypothesis that they are not drawn from the same sample and that the distributions at different distances are different. The sample of the Galactic Plane PGCCs will be extended when JPS sources are also included, and this analysis will be the subject of a further study (Eden et al., in preparation).

\subsection{Column densities of SCOPE sources}

The column densities of the SCOPE sources in the nearby environments and the distant Galactic Plane are compared. The column densities were calculated using the following:

\begin{equation}
N_{\rmn{H_{2}}} = \frac{S_{\nu, \rmn{peak}}}{B_{\nu}(T_{d})\Omega_{\rmn{b}}\kappa_{\nu}m_{\rmn{H}}\upmu}
\end{equation}

\noindent where $S_{\nu, \rmn{peak}}$ is the peak intensity, $B_{\nu}(T_{d})$ and $\kappa_{\nu}$ are as defined above, $\Omega_{b}$ is the solid angle of the beam, $m_{\rmn{H}}$ is the mass of a hydrogen atom, and $\upmu$ is the mean mass per hydrogen molecule, taken to be 2.8 \citep{Kauffmann08}.

\begin{figure}
\begin{tabular}{l}
\includegraphics[scale=0.5]{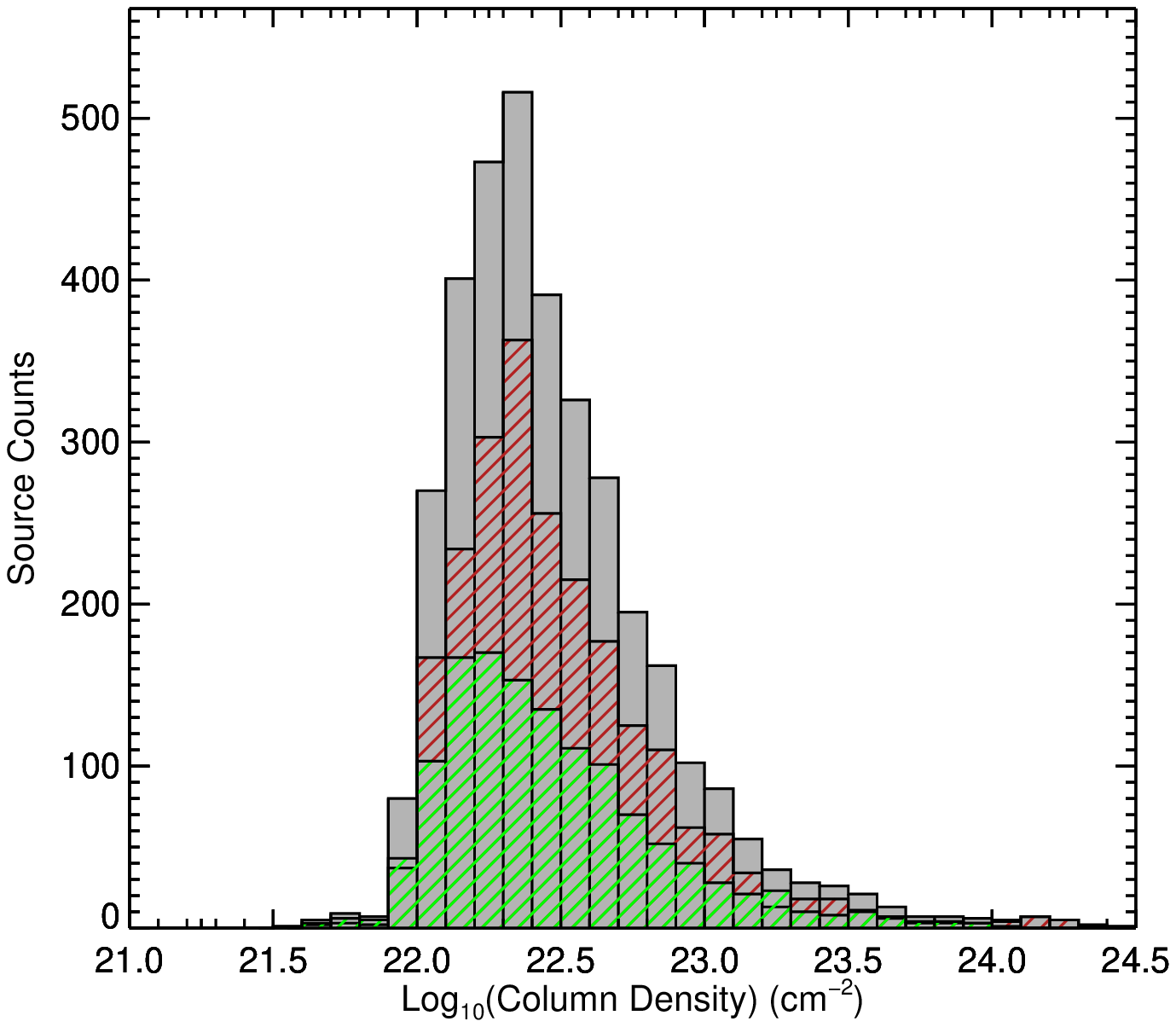} \\ \includegraphics[scale=0.5]{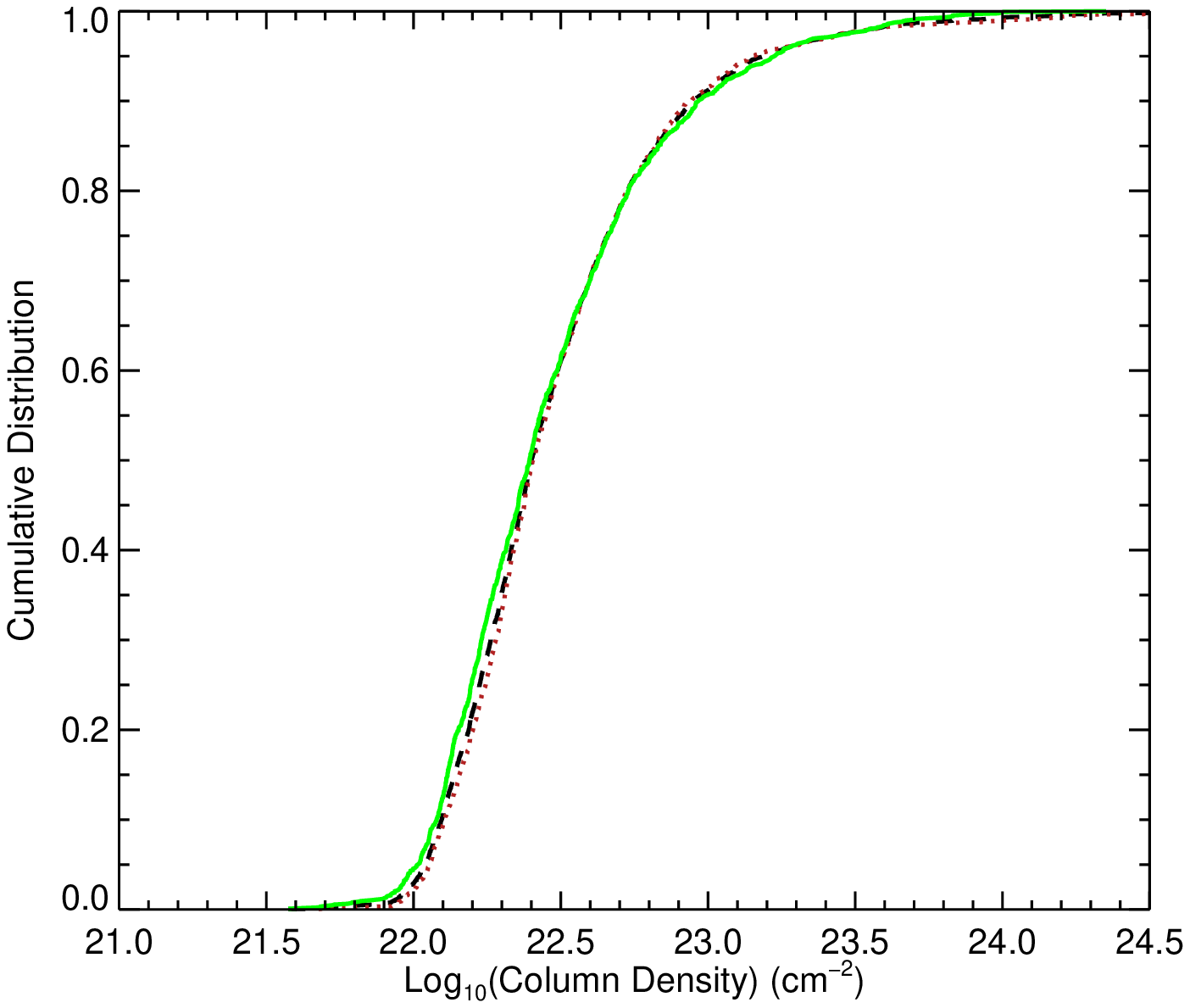}\\
\includegraphics[scale=0.5]{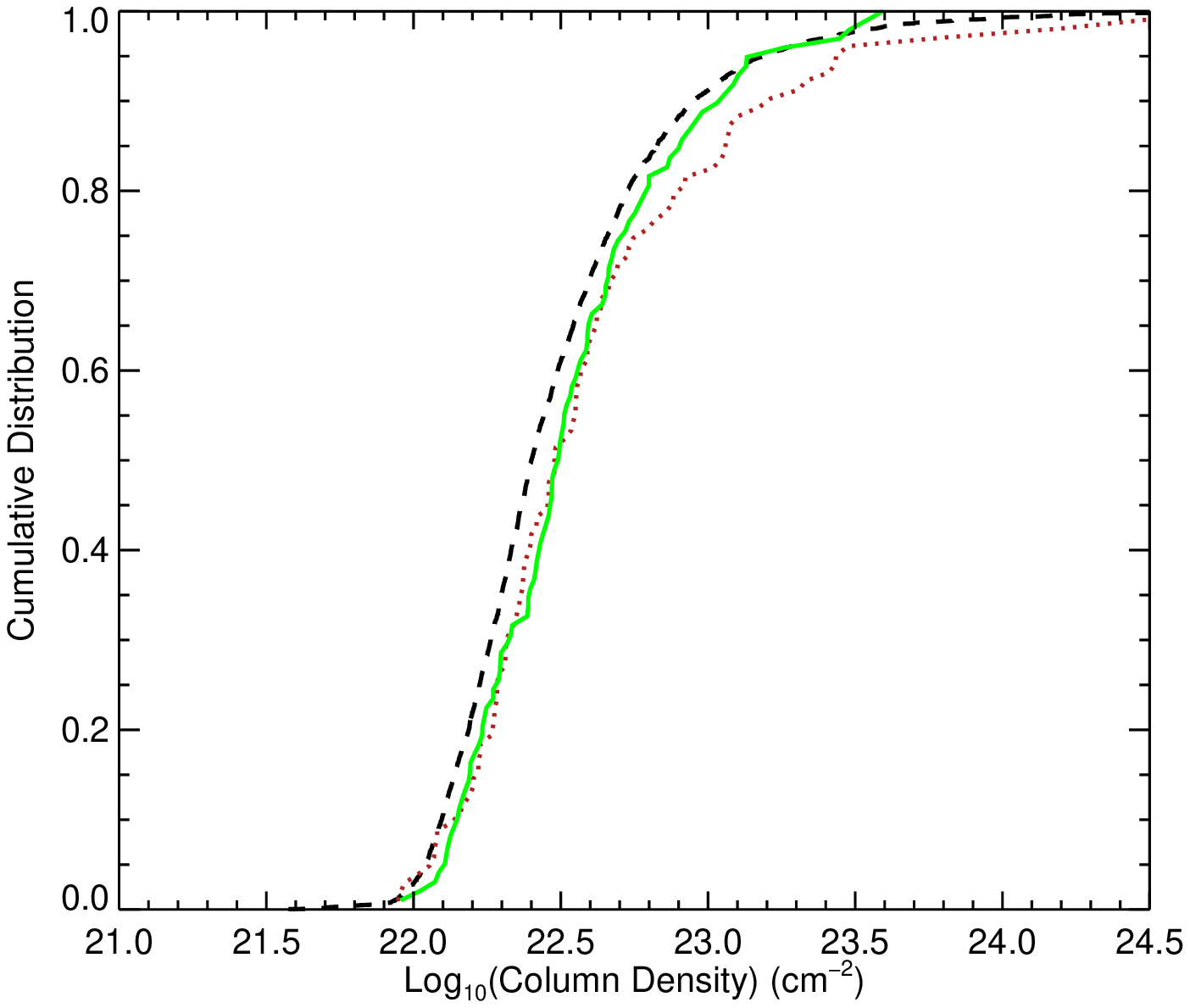}\\
\end{tabular}
\caption{Top panel: column density distribution of all SCOPE compact sources (grey histogram), with those in the distant Galactic Plane ($d\,>\,500$\,pc; red histogram) and nearby sample ($d\,<\,500$\,pc; green histogram). Middle panel: the cumulative distributions of the same histograms with the dashed black line, dotted red line, and green solid line representing, the whole sample, in the distant Galactic Plane and nearby Plane, respectively. Bottom panel: the cumulative distributions of the star-forming samples, with the dashed black line representing the total sample (as in the middle panel, includes non-star-forming sources), and the dotted red lines and green solid lines represent the star-forming sources in distant and nearby environments, respectively.}
\label{densities}
\end{figure}

The distribution of column densities for the entire SCOPE sample, as well as the two subsamples are displayed in Fig.~\ref{densities} (top, middle), with the cumulative distribution also included. A K--S test of the two subsamples gives a $\sim$\,2.5\,$\upsigma$, or a 2.5 per cent chance that they are not drawn from the same population. We therefore cannot strongly reject the null hypothesis that they are not drawn from the same sample.

When comparing only the star-forming samples, as shown in Fig.\ref{densities} (bottom), the K--S test gives a 94 per cent result that the nearby and distant Galactic Plane SCOPE sources are drawn from the same population. We can assume, however, that the total star-forming sample is considerably different from the entire SCOPE sample due to a K-S test giving a probability of their being the same of 0.001. This result is consistent with that of \citet{Urquhart14}, who found that the star-forming clumps in the ATLASGAL survey had a considerably different column-density distribution than that of the entire population.

\section{Summary}

We present the first data release of the SCOPE survey, presenting the data images and a compact source catalogue. The data consist of observations of 1235 \emph{Planck} Galactic Cold Clumps (PGCCs) at angular resolutions of 14.4 arcsec, significantly improving upon the 5-arcmin resolution of \emph{Planck}. The improved resolution reveals significant substructure within these sources, reflected by a compact source catalogue consisting of 3528 sources. The data are downloadable from the CADC, with the compact source catalogue included as Supporting Information to this article.

The compact source catalogue was produced using the {\sc FellWalker} algorithm, reflecting the same method used within the JCMT Plane Survey (JPS; \citealt{Eden17}). A comparison of peak intensities and integrated fluxes in overlapping sources between the JPS and SCOPE surveys shows slight discrepancies, but otherwise good agreement. The SCOPE sources are significantly smaller than those within the \emph{Planck} catalogue, with peaks of the angular-size distribution found at 35 arcsec compared to 8 arcmin.

The detection rate of PGCCs within the SCOPE survey is 45 per cent, with 558 PGCCs detected of the 1235 observed. The survey is 95 complete in PGCCs with column densities $N_{\rmn{H_{2}}}$\,$>$\,5\,$\times$\,10$^{21}$\,cm$^{-2}$, and to masses and luminosities of 5\,$\times$\,10$^{3}$\,M$_{\odot}$ and 1\,$\times$\,10$^{3}$\,L$_{\odot}$, respectively.

By positionally matching the SCOPE compact sources with YSOs from the WISE catalogue, and splitting the sample into sources that are within 0.5\,kpc and those at greater distances, we found that the ratio of $L/M$ is consistent between these samples. The column densities of these two samples of SCOPE sources are also consistent. The distribution of column densities of star-forming sources, however, were found to be significantly different from those of the whole SCOPE compact source catalogue.

\section*{Acknowledgements}

The authors would like to thank the anonymous referee for their comments, they have greatly improved the quality of the paper. DJE is supported by a STFC postdoctoral grant (ST/M000966/1). MJ acknowledges the support of the Academy of Finland Grant No. 285769. DC is currently supported by the Centre National d'\' {E}tudes Spatiales (CNES) through PhD grant 0102541 and the R\'{e}gion Bourgogne Franche-Comt\'{e}. MK was supported by Basic Science Research Program through the National Research Foundation of Korea (NRF) funded by the Ministry of Science, ICT \& Future Planning (No. NRF-2015R1C1A1A01052160). CWL is  supported by Basic Science Research Program through the National Research Foundation of Korea (NRF) funded by the Ministry of Education, Science and Technology (NRF-2016R1A2B4012593). JHH is supported by the NSF of China under Grant Nos. 11873086 and U1631237, partly by Yunnan province (2017HC018), and also partly by the Chinese Academy of Sciences (CAS) through a grant to the CAS South America Center for Astronomy (CASSACA) in Santiago, Chile. The James Clerk Maxwell Telescope is operated by the East Asian Observatory on behalf of The National Astronomical Observatory of Japan; Academia Sinica Institute of Astronomy and Astrophysics; the Korea Astronomy and Space Science Institute; the Operation, Maintenance and Upgrading Fund for Astronomical Telescopes and Facility Instruments, budgeted from the Ministry of Finance (MOF) of China and administrated by the Chinese Academy of Sciences (CAS), as well as the National Key R\&D Program of China (No. 2017YFA0402700). KW acknowledges support by the National Key Research and Development Program of China (2017YFA0402702), the National Science Foundation of China (11721303), and the starting grant at the Kavli Institute for Astronomy and Astrophysics, Peking University (7101502016). Additional funding support is provided by the Science and Technology Facilities Council of the United Kingdom and participating universities in the United Kingdom and Canada. The JCMT has historically been operated by the Joint Astronomy Centre on behalf of the Science and Technology Facilities Council of the United Kingdom, the National Research Council of Canada and the Netherlands Organization for Scientific Research. Additional funds for the construction of SCUBA-2 were provided by the Canada Foundation for Innovation. This research was carried out in part at the Jet Propulsion Laboratory which is operated for NASA by the California Institute of Technology. This research has made use of NASA's Astrophysics Data System. The Starlink software \citep{Currie14} is currently supported by the East Asian Observatory. DJE would like to dedicate this work to his auntie, Marjorie Eden.

\begin{landscape}
\begin{table}
\begin{center}
\caption{The SCOPE compact source catalogue. The columns are as defined in Section 4.1.}
\label{sourcecatalogue}
\begin{tabular}{llcccccccccccccc}\hline
Name & Region & RA$_{\rmn{peak}}$ & Dec$_{\rmn{peak}}$ & RA$_{\rmn{cen}}$ & Dec$_{\rmn{cen}}$ & $\upsigma_{\rmn{maj}}$ & $\upsigma_{\rmn{min}}$ & PA & $R_{\rmn{eff}}$ & $S_{\rmn{peak}}$ & $\Delta$$S_{\rmn{peak}}$ & $S_{\rmn{int}}$ & $\Delta$$S_{\rmn{int}}$ & SNR\\
& & ($^{\circ}$) & ($^{\circ}$) & ($^{\circ}$) & ($^{\circ}$) & ($\prime\prime$) & ($\prime\prime$) & ($^{\circ}$) & ($\prime\prime$) & (Jy\,arcsec$^{-2}$) & (Jy\,arcsec$^{-2}$) & (Jy) & (Jy) & \\
(1) & (2) & (3) & (4) & (5) & (6) & (7) & (8) & (9) & (10) & (11) & (12) & (13) & (14) & (15) \\
\hline
SCOPEG000.39+11.38	&	G0.49+11.38a	&	256.139	&	-22.294	&	256.139	&	-22.299	&	15	&	7	&	260	&	20	&	0.308	&	0.035	&	0.647	&	0.032	&	5.39	\\
SCOPEG000.41+11.41	&	G0.49+11.38a	&	256.120	&	-22.259	&	256.115	&	-22.258	&	17	&	11	&	207	&	29	&	0.943	&	0.025	&	4.232	&	0.212	&	16.53	\\
SCOPEG001.37+20.94	&	G001.3+20.9A1	&	248.655	&	-15.791	&	248.660	&	-15.794	&	23	&	15	&	192	&	43	&	1.132	&	0.151	&	7.936	&	0.397	&	11.26	\\
SCOPEG001.37+20.95	&	G001.3+20.9A1	&	248.647	&	-15.782	&	248.647	&	-15.781	&	21	&	18	&	132	&	46	&	1.379	&	0.202	&	14.625	&	0.731	&	13.71	\\
SCOPEG001.37+20.96	&	G001.3+20.9A1	&	248.636	&	-15.776	&	248.634	&	-15.771	&	18	&	14	&	222	&	33	&	1.038	&	0.155	&	4.771	&	0.239	&	10.33	\\
SCOPEG001.82+16.56	&	G001.8+16.5A1	&	252.559	&	-18.105	&	252.559	&	-18.104	&	24	&	15	&	190	&	42	&	0.651	&	0.081	&	4.494	&	0.225	&	7.82	\\
SCOPEG003.98+35.70	&	G4.18+35.79a	&	238.346	&	-4.808	&	238.357	&	-4.798	&	37	&	20	&	162	&	58	&	3.143	&	0.050	&	42.604	&	2.130	&	24.61	\\
SCOPEG004.00+35.71	&	G4.18+35.79a	&	238.350	&	-4.788	&	238.356	&	-4.778	&	48	&	33	&	167	&	79	&	1.515	&	0.081	&	35.849	&	1.792	&	11.86	\\
SCOPEG004.01+35.66	&	G4.18+35.79a	&	238.394	&	-4.809	&	238.390	&	-4.806	&	13	&	8	&	161	&	21	&	3.071	&	0.050	&	7.201	&	0.360	&	24.04	\\
SCOPEG004.02+35.66	&	G4.18+35.79a	&	238.398	&	-4.808	&	238.400	&	-4.801	&	15	&	10	&	140	&	26	&	2.137	&	0.027	&	6.655	&	0.333	&	16.73	\\
SCOPEG004.61+36.64	&	G004.5+36.7A1	&	237.928	&	-3.821	&	237.918	&	-3.837	&	45	&	21	&	115	&	68	&	15.535	&	0.069	&	243.776	&	12.189	&	27.44	\\
SCOPEG004.62+36.64	&	G004.5+36.7A1	&	237.927	&	-3.816	&	237.918	&	-3.817	&	29	&	8	&	156	&	37	&	13.091	&	0.062	&	68.204	&	3.410	&	23.12	\\
SCOPEG004.62+36.65	&	G004.5+36.7A1	&	237.926	&	-3.810	&	237.914	&	-3.799	&	40	&	20	&	254	&	66	&	12.707	&	0.049	&	167.376	&	8.369	&	22.44	\\
SCOPEG005.00+18.99	&	G005.0+19.0A1	&	252.391	&	-14.230	&	252.389	&	-14.229	&	9	&	4	&	143	&	11	&	1.117	&	0.026	&	0.755	&	0.038	&	17.61	\\
SCOPEG005.23-01.99	&	G5.17-1.97	&	271.305	&	-25.426	&	271.305	&	-25.427	&	14	&	10	&	195	&	26	&	0.724	&	0.077	&	2.136	&	0.107	&	7.34	\\
SCOPEG005.64+00.23	&	G5.73+0.19	&	269.408	&	-23.968	&	269.411	&	-23.972	&	22	&	13	&	219	&	36	&	1.930	&	0.147	&	9.894	&	0.495	&	10.58	\\
SCOPEG005.64+00.24	&	G5.73+0.19	&	269.395	&	-23.966	&	269.394	&	-23.962	&	25	&	17	&	239	&	51	&	12.946	&	0.153	&	62.862	&	3.143	&	71.00	\\
SCOPEG005.65+00.22	&	G5.73+0.19	&	269.416	&	-23.964	&	269.416	&	-23.960	&	28	&	14	&	254	&	47	&	1.665	&	0.225	&	10.648	&	0.532	&	9.13	\\
SCOPEG005.65+00.26	&	G5.73+0.19	&	269.381	&	-23.944	&	269.380	&	-23.944	&	8	&	6	&	123	&	14	&	0.917	&	0.050	&	1.119	&	0.056	&	5.03	\\
SCOPEG005.83$-$00.94	&	G005.91$-$01.00	&	270.615	&	-24.383	&	270.618	&	-24.382	&	15	&	9	&	227	&	26	&	1.654	&	0.050	&	5.157	&	0.258	&	12.63	\\
SCOPEG005.83$-$01.03	&	G005.91$-$01.00	&	270.704	&	-24.435	&	270.703	&	-24.443	&	31	&	14	&	97	&	35	&	0.788	&	0.040	&	3.053	&	0.153	&	6.02	\\
SCOPEG005.84$-$01.00	&	G005.91$-$01.00	&	270.676	&	-24.406	&	270.678	&	-24.412	&	32	&	24	&	267	&	65	&	2.297	&	0.246	&	22.732	&	1.137	&	17.54	\\
SCOPEG005.85$-$00.99	&	G005.91$-$01.00	&	270.682	&	-24.392	&	270.683	&	-24.390	&	15	&	15	&	102	&	35	&	1.618	&	0.254	&	8.175	&	0.409	&	12.35	\\
SCOPEG005.85$-$01.00	&	G005.91$-$01.00	&	270.693	&	-24.397	&	270.697	&	-24.399	&	18	&	14	&	204	&	32	&	0.753	&	0.105	&	3.974	&	0.199	&	5.75	\\
SCOPEG005.87$-$00.99	&	G005.91$-$01.00	&	270.693	&	-24.375	&	270.693	&	-24.378	&	28	&	12	&	243	&	40	&	1.183	&	0.164	&	8.784	&	0.439	&	9.03	\\
SCOPEG005.88$-$00.94	&	G005.91$-$01.00	&	270.648	&	-24.345	&	270.645	&	-24.352	&	27	&	25	&	109	&	58	&	0.999	&	0.138	&	15.217	&	0.761	&	7.62	\\
SCOPEG005.88$-$01.01	&	G005.91$-$01.00	&	270.709	&	-24.374	&	270.705	&	-24.372	&	22	&	13	&	257	&	38	&	2.814	&	0.397	&	14.790	&	0.739	&	21.49	\\
SCOPEG005.89$-$00.91	&	G005.91$-$01.00	&	270.624	&	-24.325	&	270.629	&	-24.320	&	23	&	14	&	161	&	39	&	0.723	&	0.091	&	5.750	&	0.287	&	5.52	\\
SCOPEG005.89$-$00.94	&	G005.91$-$01.00	&	270.656	&	-24.334	&	270.655	&	-24.331	&	15	&	12	&	240	&	32	&	1.727	&	0.193	&	5.672	&	0.284	&	13.18	\\
SCOPEG005.90$-$00.93	&	G005.91$-$01.00	&	270.650	&	-24.317	&	270.648	&	-24.315	&	16	&	10	&	234	&	31	&	1.162	&	0.123	&	4.413	&	0.221	&	8.88	\\
SCOPEG005.90$-$01.01	&	G005.91$-$01.00	&	270.721	&	-24.362	&	270.720	&	-24.371	&	46	&	24	&	261	&	72	&	1.936	&	0.248	&	36.472	&	1.824	&	14.78	\\
SCOPEG005.91$-$00.95	&	G005.91$-$01.00	&	270.675	&	-24.318	&	270.671	&	-24.319	&	25	&	18	&	256	&	51	&	2.880	&	0.294	&	14.736	&	0.737	&	21.99	\\
SCOPEG005.91$-$01.00	&	G005.91$-$01.00	&	270.724	&	-24.345	&	270.725	&	-24.344	&	20	&	10	&	141	&	30	&	1.467	&	0.354	&	9.249	&	0.462	&	11.20	\\
SCOPEG005.91$-$01.02	&	G005.91$-$01.00	&	270.742	&	-24.357	&	270.740	&	-24.354	&	23	&	21	&	223	&	54	&	2.246	&	0.442	&	20.771	&	1.039	&	17.15	\\
SCOPEG005.92$-$00.96	&	G005.91$-$01.00	&	270.683	&	-24.313	&	270.683	&	-24.311	&	12	&	9	&	117	&	22	&	0.722	&	0.121	&	1.833	&	0.092	&	5.51	\\
SCOPEG005.92$-$00.97	&	G005.91$-$01.00	&	270.696	&	-24.317	&	270.696	&	-24.316	&	13	&	12	&	170	&	29	&	0.695	&	0.138	&	2.919	&	0.146	&	5.31	\\
SCOPEG005.92$-$00.99	&	G005.91$-$01.00	&	270.714	&	-24.335	&	270.713	&	-24.335	&	31	&	23	&	156	&	62	&	2.432	&	0.550	&	27.790	&	1.389	&	18.57	\\
SCOPEG005.96$-$01.12	&	G005.91$-$01.00	&	270.857	&	-24.359	&	270.843	&	-24.361	&	23	&	15	&	162	&	36	&	0.701	&	0.032	&	4.665	&	0.233	&	5.35	\\
\hline
\multicolumn{15}{l}{$\emph{Note:}$ Only a small portion of the catalogue is shown here. The entire catalogue is available in the Supporting Information.}\\
\end{tabular}
\end{center}
\end{table}
\end{landscape}

\bibliographystyle{mnras}
\bibliography{SCOPEdescrip}

\clearpage
\onecolumn

\noindent
Author affiliations:\\
$^{1}$Astrophysics Research Institute, Liverpool John Moores University, IC2, Liverpool Science Park, 146 Brownlow Hill, Liverpool, L3 5RF, UK\\
$^{2}$Centre for Astrophysics Research, Science \& Technology Research Institute, University of Hertfordshire, College Lane, Hatfield, Herts AL10 9AB, UK\\
$^{3}$Korea Astronomy and Space Science Institute, 776 Daedeokdae-ro, Yuseong-gu, Daejon 34055, Republic of Korea\\
$^{4}$East Asian Observatory, 660 N. Aohoku Place, University Park, Hilo, Hawaii 96720, USA\\
$^{5}$Department of Physics, P.O.Box 64, FI-00014, University of Helsinki, Finland\\
$^{6}$Institute of Astronomy and Astrophysics, Academia Sinica. 11F of Astronomy-Mathematics Building, AS/NTU No.1, Sec. 4, Roosevelt Rd, Taipei 10617, Taiwan\\
$^{7}$Nobeyama Radio Observatory, National Astronomical Observatory of Japan, National Institutes of Natural Sciences, Nobeyama, Minamimaki, Minamisaku, Nagano 384-1305, Japan\\
$^{8}$NRC Herzberg Astronomy and Astrophysics, 5071 West Saanich Rd, Victoria, BC V9E 2E7, Canada\\
$^{9}$Department of Physics and Astronomy, University of Victoria, Victoria, BC V8W 2Y2, Canada\\
$^{10}$Kavli Institute for Astronomy and Astrophysics, Peking University, 5 Yiheyuan Road, Haidian District, Beijing 100871, China\\
$^{11}$Department of Astronomy, Peking University, 100871, Beijing China\\
$^{12}$Jodrell Bank Centre for Astrophysics, School of Physics and Astronomy, The University of Manchester, Oxford Road, Manchester M13 9PL, UK\\
$^{13}$National Astronomical Observatories, Chinese Academy of Sciences, Beijing, 100012, China\\
$^{14}$Key Laboratory of Radio Astronomy, Chinese Academy of Science, Nanjing 210008, China\\
$^{15}$IRAP, Universit\'{e} de Toulouse, CNRS, UPS, CNES, Toulouse, France\\
$^{16}$Max-Planck-Institut f\"{u}r Radioastronomie, Auf dem H\"{u}gel 69, 53121, Bonn, Germany\\
$^{17}$Yunnan Observatories, Chinese Academy of Sciences, 396 Yangfangwang, Guandu District, Kunming, 650216, P. R. China\\
$^{18}$Chinese Academy of Sciences South America Center for Astronomy, China-Chile Joint Center for Astronomy, Camino El Observatorio \#1515, Las Condes, Santiago, Chile\\
$^{19}$Departamento de Astronom\'{\i}a, Universidad de Chile, Camino el Observatorio 1515, Las Condes, Santiago, Chile\\
$^{20}$Department of Astronomy, Yunnan University, and Key Laboratory of Astroparticle Physics of Yunnan Province, Kunming, 650091, China\\
$^{21}$Harvard-Smithsonian Center for Astrophysics, 60 Garden Street, Cambridge, MA 02138, USA\\
$^{22}$Jet Propulsion Laboratory, California Institute of Technology, 4800 Oak Grove Drive, Pasadena, CA 91109, USA\\
$^{23}$Department of Astronomy, The University of Texas at Austin, 2515 Speedway, Stop C1400, Austin, TX 78712-1205, USA\\
$^{24}$Department of Physics and Astronomy, The Open University, Walton Hall, Milton Keynes, MK7 6AA, UK\\
$^{25}$RAL Space, STFC Rutherford Appleton Laboratory, Chilton, Didcot, Oxfordshire, OX11 0QX, UK\\
$^{26}$Korea University of Science \& Technology, 176 Gajeong-dong, Yuseong-gu, Daejeon, Republic of Korea\\
$^{27}$E\"{o}tv\"{o}s Lor\'{a}nd University, Department of Astronomy, P\'{a}zm\'{a}ny P\'{e}ter s\'{e}t\'{a}ny 1/A, H-1117, Budapest, Hungary\\
$^{28}$Konkoly Observatory of the Hungarian Academy of Sciences, H-1121 Budapest, Kinkily Thege Mikl\'{o}s\'{u}t 15-17, Hungary\\
$^{29}$School of Space Research, Kyung Hee University, Yongin-Si, Gyeonggi-Do 17104, Korea\\
$^{30}$SOFIA Science Center, Universities Space Research Association, NASA Ames Research Center, Moffett Field, California 94035, USA\\
$^{31}$School of Physics \& Astronomy, Cardiff University, Queens Buildings, The Parade, Cardiff, CF24 3AA, UK\\
$^{32}$Institute of Astronomy, National Central University, Jhongli 32001, Taiwan\\
$^{33}$Physical Research Laboratory, Navrangpura, Ahmedabad, Gujarat 380009, India\\
$^{34}$Department of Physics and Astronomy, University of Waterloo, Waterloo, Ontario, N2L 3G1, Canada\\
$^{35}$Institute of Physics I, University of Cologne, Z\"{u}lpicher Str. 77, D-50937, Cologne, Germany\\
$^{36}$European Space Astronomy Centre (ESA/ESAC), Operations Department, Villanueva de la Ca\~{n}ada(Madrid), Spain\\
$^{37}$Department of Physics, The Chinese University of Hong Kong, Shatin, NT, Hong Kong SAR\\
$^{38}$Departamento de Astronom\'ia, Universidad de Concepci\'on, Av. Esteban Iturra s/n, Distrito Universitario, 160-C, Chile\\
$^{39}$Chinese Academy of Sciences South America Center for Astronomy\\
$^{40}$Astronomy Department, University of California, Berkeley, CA 94720, USA\\
$^{41}$Department of Physics and Astronomy, University College London, Gower Street, London, WC1E 6BT, UK\\
$^{42}$Max-Planck-Institut f\"{u}r Extraterrestrische Physik, Giessenbachstrasse 1, 85748, Garching, Germany\\
$^{43}$Xinjiang Astronomical Observatory, Chinese Academy of Sciences; University of the Chinese Academy of Sciences\\
$^{44}$ESA/STScI, 3700 San Martin Drive, Baltimore, MD 21218 USA\\
$^{45}$Department of Physics, National Tsing Hua University, Guangfu Road, East District, Hsinchu City, Taiwan\\
$^{46}$Institut UTINAM, UMR 6213, CNRS, Univ Bourgogne Franche Comte, France\\
$^{47}$Department of Physical Science, Graduate School of Science, Osaka Prefecture University, 1-1 Gakuen-cho, Naka-ku, Sakai, Osaka 599-8531, Japan\\
$^{48}$Chile Observatory, National Astronomical Observatory of Japan, National Institutes of Natural Science, 2-21-1 Osawa, Mitaka, Tokyo 181-8588, Japan\\
$^{49}$Center for Computational Sciences, The University of Tsukuba, 1-1-1, Tennodai, Tsukuba, Ibaraki 305-8577, Japan\\
$^{50}$School of Physics, University of New South Wales, Sydney, NSW 2052, Australia\\
$^{51}$Department of Physics, School of Science and Technology, Nazarbayev University, Astana 010000, Kazakhstan\\
$^{52}$University of Leicester, Physics \& Astronomy, 1 University Road, Leicester LE1 7RH, UK\\
$^{53}$Department of Physics and Atmospheric Science, Dalhousie University, Halifax, NS, B3H 4R2, Canada\\
$^{54}$Institute of Astronomy and Department of Physics, National Tsing Hua University, Hsinchu, Taiwan\\
$^{55}$Key Laboratory for Research in Galaxies, and Cosmology, Shanghai Astronomical Observatory, Chinese Academy of Sciences, 80 Nandan Road, Shanghai 200030, China\\
$^{56}$Purple Mountain Observatory, Chinese Academy of Sciences, China, Nanjing 210008\\
$^{57}$Square Kilometre Array Organisation, Jodrell Bank Observatory, Lower Withington, Macclesfield, Cheshire SK11 9DL, UK\\
$^{58}$Department of Earth Science and Astronomy, Graduate School of Arts and Sciences, The University of Tokyo, 3-8-1 Komaba, Meguro, Tokyo 153-8902, Japan\\
$^{59}$LERMA, Observatoire de Paris, PSL Research University, CNRS, Sorbonne Universit\'es, UPMC Univ. Paris 06, Ecole normale sup\'erieure, 75005 Paris, France\\
$^{60}$Konkoly Observatory, Research Centre for Astronomy and Earth Sciences, Hungarian Academy of Sciences, H-1121 Budapest, Konkoly Thege Mikl\'{o}s \'{u}t 15-17, Hungary\\
$^{61}$Physikalisches Institut, Universit\"{a}t zu K\"{o}ln, Z\"{u}lpicher Str. 77, D-50937 K\"{o}ln, Germany\\
$^{62}$CSIRO Astronomy and Space Science, PO Box 76, Epping NSW, Australia\\
$^{63}$Xinjiang Astronomical Observatory, Chinese Academy of Sciences; University of the Chinese Academy of Sciences\\
$^{64}$Astronomy Department, Abdulaziz University, PO Box 80203, 21589, Jeddah, Saudi Arabia\\
$^{65}$UK Astronomy Technology Center, Royal Observatory, Blackford Hill, Edinburgh EH9 3HJ, UK; Institute for Astronomy, University of Edinburgh, Royal Observatory, Blackford Hill, Edinburgh EH9 3HJ, UK\\
$^{66}$School of Physics and Astronomy, Queen Mary University of London, Mile End Road, London E1 4NS, UK\\
$^{67}$Hiroshima Astrophysical Science Center, Hiroshima University, Kagamiyama, Higashi-Hiroshima, Hiroshima 739-8526, Japan\\
$^{68}$Department of Physical Science, Hiroshima University, Kagamiyama, Higashi-Hiroshima 739-8526, Japan\\
$^{69}$Department of Physics and Astronomy, Seoul National University, Gwanak-gu, Seoul 08826, Korea\\
$^{70}$Korea University of Science and Technology, 217 Gajeong-ro, Yuseong-gu, Daejeon 34113, Republic of Korea\\
$^{71}$Department of Earth Sciences, National Taiwan Normal University, 88 Sec. 4, Ting-Chou Road, Taipei 116, Taiwan\\
$^{72}$Department of Physics and Astronomy, McMaster University, Hamilton, ON L8S 4M1 Canada\\
$^{73}$Department of Physics, The Chinese University of Hong Kong, Shatin, New Territory, Hong Kong, China\\
$^{74}$Infrared Processing Analysis Center, California Institute of Technology, 770 South Wilson Ave., Pasadena, CA 91125, USA\\
$^{75}$School of Astronomy and Space Science, Nanjing University, Nanjing 210023, China\\
$^{76}$Astrophysics Group, Cavendish Laboratory, J J Thomson Avenue, Cambridge CB3 0HE, UK\\
$^{77}$Graduate School of Informatics and Engineering, The University of Electro-Communications, Chofu, Tokyo 182-8585, Japan\\
$^{78}$Department of Physics and Astronomy, Graduate School of Science and Engineering, Kagoshima University, 1-21-35 Korimoto, Kagoshima 890-0065, Japan\\
$^{79}$Jeremiah Horrocks Institute for Mathematics, Physics \& Astronomy, University of Central Lancashire, Preston PR1 2HE, UK\\
$^{80}$IAPS-INAF, via Fosso del Cavaliere 100, 00133, Rome, Italy\\
$^{81}$International Centre for Radio Astronomy Research, Curtin University, GPO Box U1987, Perth, WA 6845, Australia\\
$^{82}$Aix Marseille Universit\'e, CNRS, LAM (Laboratoire d'Astrophysique de Marseille) UMR 7326, F-13388, Marseille, France\\

\bsp
\label{lastpage}

\end{document}